\documentclass[sigconf, arxiv]{acmart}

\AtBeginDocument{%
  \providecommand\BibTeX{{%
    \normalfont B\kern-0.5em{\scshape i\kern-0.25em b}\kern-0.8em\TeX}}}

\usepackage{geometry}
\usepackage{multirow}
\usepackage{amsfonts} 

\usepackage{url}            
\usepackage{booktabs}       
\usepackage{amsfonts}       
\usepackage{nicefrac}       
\usepackage{microtype}      
\usepackage{lipsum}
\usepackage{bbm}
\usepackage{textcomp}
\usepackage{xcolor}
\usepackage[utf8]{inputenc}
\usepackage[english]{babel}
\usepackage{amsthm}
\usepackage{amsmath}
\usepackage{microtype}
\usepackage{booktabs}
\usepackage{chngpage}
\usepackage{subcaption}
\usepackage{xr}
\usepackage{dsfont}
\usepackage{algorithm,algpseudocode}
\usepackage{hyperref}
\usepackage{wrapfig}
\usepackage{titlesec}
\usepackage[export]{adjustbox} 

\def\environment{\mathcal{E}}

\def\actions{\mathcal{A}}

\def\observations{\mathcal{O}}

\def\E{\mathbb{E}}

\def\Pr{\mathbb{P}}

\def\1{\mathbf{1}}

\DeclareMathOperator*{\argmax}{arg\,max}
\DeclareMathOperator*{\argmin}{arg\,min}

\theoremstyle{thmstyleone}%
\theoremstyle{thmstyletwo}%

\theoremstyle{thmstylethree}%

\theoremstyle{definition}
\newtheorem{exmp}{Example}

\copyrightyear{2023}
\acmYear{2023}
\setcopyright{acmlicensed}
\acmConference[RecSys '23]{Seventeenth ACM Conference on Recommender Systems}{September 18--22, 2023}{Singapore, Singapore}
\acmBooktitle{Seventeenth ACM Conference on Recommender Systems (RecSys '23), September 18--22, 2023, Singapore, Singapore}
\acmPrice{15.00}
\acmDOI{10.1145/3604915.3608855}
\acmISBN{979-8-4007-0241-9/23/09}

\begin{document}

\title[Deep Exploration for Recommendation Systems]{Deep Exploration for Recommendation Systems}

\author{Zheqing Zhu}
\email{billzhu@fb.com}
\affiliation{%
  \institution{Meta AI, Stanford University}
  \city{Menlo Park}
  \state{CA}
  \country{USA}
}

\author{Benjamin Van Roy}
\email{bvr@stanford.edu}
\affiliation{%
  \institution{Stanford University}
  \city{Stanford}
  \state{CA}
  \country{USA}
}

\renewcommand{\shortauthors}{Zhu and Van Roy}

\begin{abstract}
  Modern recommendation systems ought to benefit by probing for and learning from delayed feedback.  Research has tended to focus on learning from a user's response to a single recommendation.  Such work, which leverages methods of supervised and bandit learning, forgoes learning from the user's subsequent behavior.  Where past work has aimed to learn from subsequent behavior, there has been a lack of effective methods for probing to elicit informative delayed feedback.  Effective exploration through probing for delayed feedback becomes particularly challenging when rewards are sparse.  To address this, we develop deep exploration methods for recommendation systems.  In particular, we formulate recommendation as a sequential decision problem and demonstrate benefits of deep exploration over single-step exploration. Our experiments are carried out with high-fidelity industrial-grade simulators and establish large improvements over existing algorithms.
\end{abstract}

\maketitle
\section{Introduction}
Recommendation systems (RS) play a critical role in helping users find desired content within the ever-expanding universe of information that can be accessed via the Internet.  Supervised learning procedures, such as collaborative filtering\cite{schafer2007collaborative} and content-based filtering\cite{blanda2016online} that form the cornerstone of RS, are typically used to model the probability that each piece of content will immediately engage a given user. However, those methods are inadequate when optimizing for delayed user feedback \cite{joulani2013online, ktena2019addressing}. In particular, the value of a recommendation can become evident in later interactions with a user, rather than through immediate engagement. For example, a user might submit a positive rating about a news platform after a series of recommendations.  

Delays in relevant feedback are prevalent in practical RS. One approach to developing a RS that accounts for delayed feedback is to maintain a model that predicts cumulative future value from interactions with each user, using reinforcement learning (RL) algorithms \cite{liu2005integrating,shih2008product,desirena2019maximizing,iwata2008recommendation, eugene2019slateq, theocharous2015personalized, chen2019top, zheng2018drn}. Maximizing cumulative future value is especially important when making recommendations that can trigger informative feedback only after subsequent interactions with the user. Although the aforementioned work has demonstrated potential value of RL in RS, sparse and delayed feedback poses challenges to fully realizing the value of RL.

A major challenge to existing RS algorithms arises when positive feedback is rare and tends to arise only after a series of suitable recommendations. For the aforementioned methods to learn from such feedback, enormous quantities of data are required.  Indeed, data requirements grow proportionately with the ratio between uninformative and informative feedback \cite{ganganwar2012overview}.

\begin{figure}[t]
     \centering
     \includegraphics[width=0.3\textwidth]{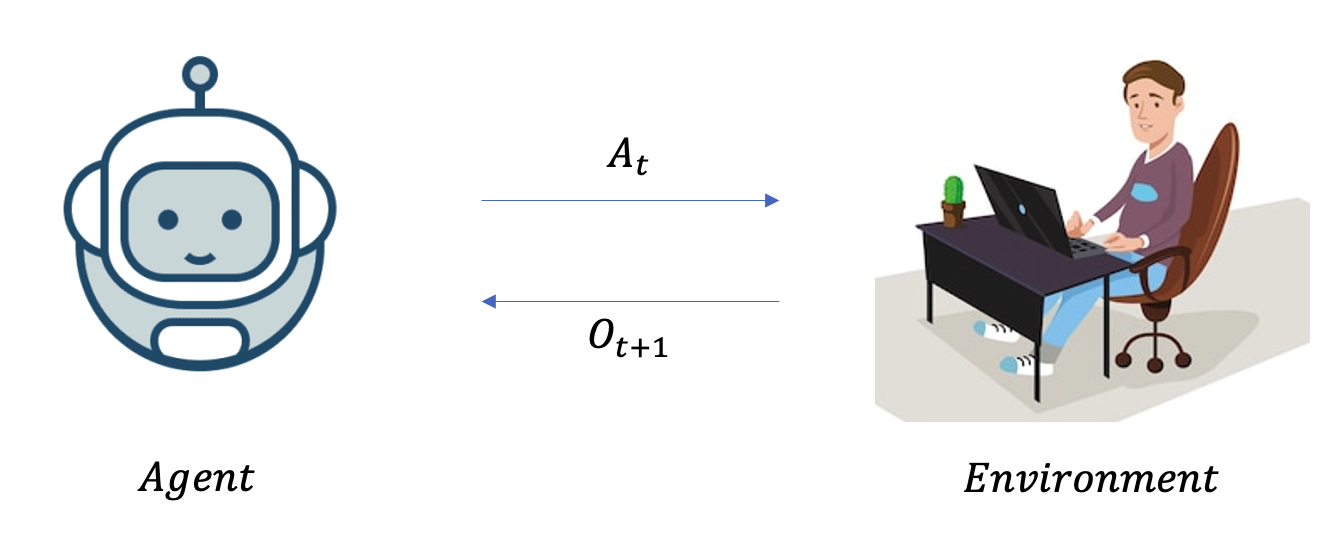}
     \caption{Real-World RS Interface.}
     \label{fig:interface}
\end{figure}

Recommendations that aim to elicit informative feedback can dramatically reduce data requirements.  
Some recent research focuses on bandit-based exploration algorithms such as upper-confidence bound (UCB) \cite{nguyen2019recommendation} and Thompson sampling \cite{chapelle2011empirical} algorithms.  However, such exploration strategies are myopic in the sense that they seek information only from immediate user feedback.  Deep exploration \cite{osband2019deep} algorithms, on the other hand, further seek information revealed by sequences of rather than individual recommendations.  This enables RS to more quickly learn users' preferences even when their feedback is sparse and delayed.

To the best of our knowledge, this paper is the first to demonstrate the importance of deep exploration in RS.  Our work also points to potential benefits beyond user personalization.  We develop the first deep exploration algorithm for RS, leveraging randomized value functions \cite{osband2019deep} and epistemic neural networks (ENNs) \cite{osband2021epistemic}.  This leads to a proof of concept and empirical evidence that deep exploration can dramatically increase cumulative positive feedback.  In Section 2, we review research on RL and bandit learning for RS.  In Section 3, we formulate RS design in terms of a sequential decision problem.  In Section 4, we present results from case studies that compare common RS designs and bandit exploration designs against one that leverages deep exploration. In doing so, we demonstrate benefits of deep exploration in the face of sparse and delayed feedback. In Section 5, we offer a brief introduction to ENNs.  In Section 6, we present a scalable deep exploration algorithm based on randomized value functions and ENNs. In Section 7, we study the effectiveness of deep exploration in RS through a set of experiments, including a toy environment, a high-fidelity e-commerce environment, and a slate ranking environment with parallel user streams.  In Section 9, we consider implications of our empirical results, highlighting potential opportunities and possible steps toward productionization.

\section{Related Work}
In this section, we cover a two topics in RS that are related to sparse and delayed feedback, Contextual Bandit and Exploration as well as Reinforcement Learning.
\vspace{-0.05in}
\subsection{Contextual Bandit and Exploration}\label{sec:bandit}
Contextual bandit algorithms are core to RS research for personalization. Contextual bandit \cite{langford2007epoch} is a problem built upon Multi-armed Bandit \cite{katehakis1987multi} where each bandit is contextualized with a representation such that algorithms can make inference across different contexts. In a contextual bandit problem, an agent is asked to select an arm based on the bandit's context. The agent then uses the collected feedback from the selected arm to update its posterior belief over the bandit. Contextual bandits have been well adopted in the RS community with applications in news article recommendation \cite{li2010contextual, li2011unbiased}, mobile recommendations \cite{bouneffouf2012contextual}, and even virtual reality \cite{heyse2019contextual}.

One key advantage of adopting contextual bandit is to leverage the mechanism of exploration. Exploration is a strategic behavior where an agent seeks for information rather than optimizing for reward accumulation. There are multiple strategies to carry out exploration in a contextual bandit environment. The most common exploration strategy in RS is $\epsilon$-greedy \cite{kamishima2011personalized, theocharous2015personalized, hu2017playlist}. $\epsilon$-greedy keeps a point estimate for the value of each recommendation and provides a random recommendation with probability $\epsilon$, which may or may not help learn more information about this recommendation. Beyond $\epsilon$-greedy, an agent can adopt the upper confidence bound (UCB) \cite{garivier2011kl, auer2002finite, Guo2023evaluate} strategy, where the agent follows the "optimism in the face of uncertainty" principle. An agent deems the upper bound of the reward of an arm as its belief of the environment and updates the confidence bound of each arm as it interacts with the environment. The strategy is extended to the LinUCB algorithm \cite{li2010contextual} for logistic regression modeling and neural networks \cite{zhou2019neural}. Another branch of exploration strategy follows Thompson sampling \cite{thompson1933likelihood}, where an agent samples from its posterior belief of bandit arms' rewards and selects the arm with highest sample estimate. Empirical results also show that Thompson sampling outperforms $\epsilon$-Greedy in cumulative positive feedback when only optimizing for immediate user feedback \cite{chapelle2011empirical}. The algorithm is also extended to logistic regression \cite{chapelle2011empirical} and neural network versions through ensemble sampling \cite{lu2017ensemble, QinEnsembleSampling}, neural Thompson sampling \cite{zhang2020neural,OsbandNeuralTestbed} or epistemic neural network powered Thompson sampling \cite{zhu2023scalable}. However, both UCB-style algorithm and Thompson-sampling-style algorithms for Contextual Bandits remain to explore myopically, i.e. only explore for immediate reward. 
\subsection{Reinforcement Learning}
Recent research in RS has expanded its focus beyond contextual bandits and starts to study the possible design of reinforcement learning algorithms in RS. A key advantage of reinforcement learning over contextual bandit algorithms is that it optimizes beyond immediate reward and maximizes cumulative reward. Using reinforcement learning, we have seen significant user satisfaction improvements with massive dataset available for learning from offline reinforcement learning with value function approximation \cite{chen2019generative, rojanavasu2005new, tang2019reinforcement, golovin2004reinforcement, xu2023optimize}, policy gradient \cite{chen2019top} and actor-critic \cite{chen2022off}, but none of the work above focuses on the data efficiency of their solutions and consumes a massive amount of data to produce a reinforcement learning model. Exploration in deep reinforcement learning goes beyond the exploration strategy in bandits. Agents leveraging reinforcement learning aims to optimize long-term cumulative reward and hence needs to seek for information regarding long-term cumulative reward of a policy instead of immediate reward. There has been some recent work \cite{chen2021values} that starts to focus on the possibility of exploration for users in RS but the design remains to either tweak reward function to incentivize exploration or change the loss function and representation to encode exploration, which only facilitates myopic exploration, where as a deep exploration agent should be able to explore multiple step ahead in a strategic fashion. 

In the next two sections, we will define RS in the context of sequential decision making and cover case studies to illustrate the benefit of deep exploration in an intuitive fashion.
\section{Problem Formulation}\label{prob}
In this section, we formulate RS as a sequential decision problem in which the agent interacts with a community of users.  Each user is assigned a distinct positive integer index.  At each time $t$, there is a set $\mathcal{U}_t$ of active users.  Any particular user may be in any combination of these sets across time.  At each time, the agent registers an observation from and then provides a recommendation to each active user.

We model interactions as generated by an environment, which is identified by a quintuple $\mathcal{E} = (\mathcal{O}, \mathcal{A}, r, c, \rho)$: 
\begin{enumerate}
\item Observation set $\observations$:  At each time $t$, the agent receives an observation $O_{t,u}$ from each active user $u \in \mathcal{U}_t$.  The set $\observations$ is made up of all possible values of $O_{t,u}$.
\item Action set $\mathcal{A}$: At each time $t$, the agent applies an action $A_{t,u}$ to each active user $u \in \mathcal{U}_t$.  The set $\actions$ is made up of all possible values of $A_{t,u}$.
\item Reward function $r$:  The reward function generates a real-valued reward $R_{t,u} = r(O_{t,u})$ based on the observation $O_{t,u}$.
\item Constraint function $c$: It is common in RS that not all items are available to all users at all times.  We model this in terms of a constraint function $c$, which prescribes a set of allowed recommendations.  In particular, given an observation $O_{t,u}$, the corresponding action is constrained so that $A_{t,u} \in \mathcal{A}_{t,u} = c(O_{t,u}) \subseteq \actions$. 
\item Observation probability function $\rho$: Let $O_t = \{(u, O_{t,u}) \mid u \in \mathcal{U}_t\}$ encode all the agent observes at time $t$, and let $A_t = \{(u, A_{t,u}) \mid u \in \mathcal{U}_t\}$ encode all actions executed by the agent.  Let $H_t = (O_0, A_0, O_1, \ldots, O_t) \in \mathcal{H}$ denote the history of the agent's experience through time $t$.  Given $H_t$ and $A_t$, the observation probability function generates a distribution $\rho(\cdot \mid H_t,A_t) = \Pr(O_{t+1} = \cdot \mid H_t,A_t)$.
\end{enumerate}

Agents we will develop address environments of the form we have described.  The behavior of an agent can be characterized by a function $\pi_{\rm agent}$, which, for any history $h$, specifies a probability mass function $\pi_{\rm agent}(\cdot\mid h)$ over actions.  The agent selects each action according to $A_t \sim \pi_{\rm agent}(\cdot\mid H_t)$.  To assess the performance of an agent in a environment $\environment = (\mathcal{O}, \mathcal{A}, r, c, \rho)$ over $T$ timesteps, we compute its cumulative reward
$$\sum_{t=0}^{T-1} \sum_{u \in \mathcal{U}_{t+1}} R_{t+1,u}.$$ 
Note that this formulation is very general; interactions arising in just about any recommendation system can be modeled in these terms.

In our computational experiments, we posit particular environments and simulate interactions.  Realized rewards depend on simulated randomness and thus the initial random seed.  To reduce dependence on this seed, we average results over many independent simulations, aiming to compute a close approximation to the expectation over this uncertainty:
$$\E_{\pi_{\rm agent}}\left[\sum_{t=0}^{T-1} \sum_{u \in \mathcal{U}_{t+1}} R_{t+1,u} \right].$$ 
The subscript $\pi_{\rm agent}$ indicates how actions are selected. 

We offer a simple, toy example that serves to crystallize our environment formalism.

\begin{exmp} \textbf{Toy Environment} \label{exmp:toy}\\
Consider an environment with a single user.  In particular, $\mathcal{U}_t = \{1\}$ for all $t$.  Suppose that, at each time, the agent has a single item at hand -- say, a news article -- and can choose to recommend it to the user if the user is engaged.  No item is available at more than one point in time.  The observation set is $\observations = \{0,1\}$, with $O_{t, 1}=1$ indicating that the user is engaged.  The action set is $\actions = \{0,1\}$, where $A_{t, 1}=1$ indicates recommending the item at hand.  The constraint function is defined by $c(1) = \actions$ and $c(0) = \{0\}$ to express that the item can only be recommended when the user is engaged.  The reward function is defined by $r(o) = o$, indicating a preference for engagement.  The observation probability function determines the chances that the user will be engaged in the next timestep.  

To offer a concrete, albeit contrived, example, suppose the user remains engaged until receiving ten recommendations over consecutive timesteps, at which points the user is triggered to disengage for one hundred timesteps.  We let $Y_t$ indicate whether the user is in this disengaged state due to being triggered within the past hundred timesteps.  The observation probability function is then defined by
$$\rho((1,1) \mid H_t,A_t) = \left\{\begin{array}{ll}
0 & \text{if } Y_t = 1 \\
1 & \text{if } Y_t = 0.
\end{array}\right.$$
In this environment, to maximize cumulative reward over a long horizon, the agent ought to avoid recommending every tenth item but recommend each item between those omissions.
\end{exmp}

In the next section, we will offer another example environment that embodies sparse and delayed reward and a couple of analyses to explain the benefits of deep exploration.
\section{Why Deep Exploration?}\label{why}

Exploration is vital to sample efficiency when faced with sparse and delayed user feedback. Past research on RS has addressed exploration through either $\epsilon$-greedy schemes \cite{kamishima2011personalized, theocharous2015personalized, hu2017playlist} in a reinforcement learning context, or upper confidence bound (UCB) \cite{nguyen2019recommendation, nakamura2015ucb} and Thompson sampling \cite{chapelle2011empirical, broden2018ensemble} in a bandit learning context. The first approach randomly perturbs the exploitation strategy without strategically seeking useful information. The latter two approaches are guided by epistemic uncertainty of immediate feedback and lead to myopic exploration. However, actions of an intelligent agent ought to also depend on epistemic uncertainty about delayed feedback.  We consider an approach that addresses this via estimating epistemic uncertainty about the optimal value function. Here, the optimal value function of a history-action pair $(h, a)$, denoted by $Q : \mathcal{H} \times \actions \rightarrow \mathbb{R}$, refers to the expected reward an agent can accumulate by taking action $a$ at the history $h$ and adhering to an optimal policy thereafter.

\textbf{Deep exploration} refers to a strategic approach that positions an agent to more effectively gather information over subsequent time steps \cite{osband2019deep}. In this section, we investigate a deep exploration agent that maintains an approximation to the posterior distribution of the optimal value functions, samples a value function from this distribution upon engaging with a user and adheres to that sample until the user disengages. We refer to the period of user engagement as the user life-cycle. To evaluate the effectiveness of such an agent, we introduce SeqRec, which provides an example environment featuring sparse and delayed rewards. SeqRec represents a RS scenario in which users do not provide any feedback to the agent until the agent's sequence of recommendations successfully meets their needs.

\begin{exmp} \label{exmp:case_study}
\textbf{SeqRec Environment}\\
The SeqRec environment features a fixed set of users across all time steps, such that $\mathcal{U}_t = \mathcal{U} = \{1, 2, \dots, N\}$ for all $t$. More specifically, the SeqRec environment is defined by the following components, instantiating the environment quintuple $(\observations, \actions, r, c, \rho)$:
\begin{enumerate}
    \item The observation set $\observations = \{0, 1\}^2$. $O_{t,u} = (\text{Satisfied}_{t,u}, \text{Leave}_{t,u}) \in \observations$, where $\text{Satisfied}_{t, u} \in \{0, 1\}$ indicates whether user $u$ is satisfied with the recommendation, and $\text{Leave}_{t, u} \in \{0, 1\}$ indicates whether the user $u$ intends to disengage with the RS agent. $O_{0} = \{((u, (0, 0)) \mid u \in \mathcal{U}_0\}$.
    \item The action set $\actions = \{a_1, a_2, \text{no-op}\}$, where $a_1$ and $a_2$ are the only two genres of content for recommendation in the environment. 
    \item Reward $R_{t, u} = r(O_{t, u}) = \text{Satisfied}_{t, u}$.
    \item The available set of actions $\mathcal{A}_{t, u}$ at time step $t$ for user $u$ and the constraint function $c$ are defined as 
    $$
        \mathcal{A}_{t, u} = c(O_{t, u}) = 
        \begin{cases}
            \{a_1, a_2\}, & \text{if } \text{Leave}_{t, u} = 0,\\
            \{\text{no-op}\}, & \text{if } \text{Leave}_{t, u} = 1.
        \end{cases}
    $$
    \item The observation probability mass function $\rho$ is described by the following deterministic mechanism. 
    \begin{enumerate}
        \item The environment tracks user satisfaction level $Y_{t, u} \in \mathbb{R}$ for every user $u \in \mathcal{U}_t$. The environment uses an internal function, $g : \mathcal{U} \times \mathcal{A} \rightarrow \mathbb{R}$, to calculate the change in user satisfaction given an action. 
        $$
        Y_{t+1, u} = \begin{cases}
            Y_{t, u} + g(u, A_{t, u}), &\text{if } \text{Leave}_{t, u} = 0,\\
            0, &\text{if } \text{Leave}_{t, u} = 1,
        \end{cases}
        $$
        \item The environment tracks each user $u$'s life-cycle length at time $t$ as $L_{t, u}$. 
        $$
        L_{t+1, u}  = \begin{cases}
            L_{t, u} + 1, &\text{if } \text{Leave}_{t, u} = 0,\\
            0, &\text{if } \text{Leave}_{t, u} = 1,
        \end{cases}
        $$
        \item Each user $u$ has a target satisfaction level $b_u$, where \\
        $\text{Satisfied}_{t+1, u} = 1$ if $Y_{t+1, u} > b_u$.
        \item A user decides to disengage with the agent, $\text{Leave}_{t+1, u} = 1$, when the user is satisfied with the agent, $Y_{t+1, u} \geq b_u$, the user's satisfaction level drops below 0, $Y_{t+1, u} < 0$, or the user's current life-cycle with the agent is longer than their engagement time budget $L_{t+1, u} >= \tau_u, \tau_u \in \mathbb{N}$. 
        \item Lastly we have the probability mass function $\rho$ as
        $$
        \rho\big(O_{t+1} = \{(u, (\text{Satisfied}_{t+1, u}, \text{Leave}_{t+1, u})) \mid u \in \mathcal{U}_{t+1}\} \mid H_t, A_t \big) = 1.
        $$
    \end{enumerate}
\end{enumerate}
\end{exmp}

SeqRec represents an abstract and simplified model of user interaction. Delayed user feedback in this model is driven by the latent random variable $Y_{t,u}$ that tracks a users level of satisfaction.  The agent receives positive feedback when the user's satisfaction crosses a threshold.  Delayed feedback is commonly observed in practical RSs arising in ratings on software applications, cumulative revenue, and total engagement time.

In the following analyses, we will compare deep exploration against $\epsilon$-greedy, Thompson sampling with Gaussian priors, and UCB in a set of case studies using SeqRec. 

\subsection{SeqRec Sample Complexity: Single-User Case} \label{example-1}

\textit{Consider a SeqRec environment with a single user, $\mathcal{U} = \{1\}$, with a satisfaction change function $g(1, a_1) = 0$ and $g(1, a_2) = 1$. The user's satisfaction target level and engagement time budget are $b_1 = \tau_1 = T$. The agent's goal is to maximize cumulative reward in SeqRec. The agent has prior knowledge that only one of $a_1$ and $a_2$ increases the user's satisfaction level.}

The optimal approach here is to consistently apply either action $a_1$ or $a_2$ until the user disengages, in order to determine which action leads to an increased satisfaction level. If the optimal strategy is not employed, the agent will not receive any positive rewards before the user disengages. In the following analysis, we demonstrate the difference between $\epsilon$-greedy, myopic Thompson sampling, myopic UCB and deep exploration.

Since the agent cannot observe any differences until it applies action $a_2$ for $T$ times, there is a unique sequence of recommendations that will reveal the optimal action. Employing the $\epsilon$-greedy strategy, the expected number of user life-cycles required to learn the optimal policy is $\Theta(2^T)$, as there are $2^{T}$ possible permutations of recommendation sequences in $T$ steps. 

Consider a myopic Thompson sampling bandit learning agent with zero-mean Gaussian priors. Given a history $h$, the agent samples from the posterior distributions over the immediate reward and selects the action with the higher reward sample. Since both actions' immediate reward posteriors are zero-mean Gaussians before receiving any positive feedback, there is a $1/2$ probability of choosing either action $a_1$ or $a_2$. This probability remains constant after each posterior update until the desired sequence, applying action $a_2$ for $T$ times, is discovered. Thus, the expected number of user life-cycles with Thompson sampling is also $\Theta(2^T)$.

For the myopic UCB strategy, a common approach uses $\sqrt{\frac{\log(t)}{N_t(a)}}$ as the exploration bonus, in addition to the marginal reward estimate for a history-action pair, where $N_t(a)$ denotes the number of times action $a$ is executed up to timestep $t$. When the user starts to engage, the agent is equally optimistic about $a_1$ and $a_2$, with identical reward estimates. After each step, the agent reduces its exploration bonus for the selected action, ensuring that the same action will not be chosen in the next timestep. Consequently, the agent continuously alternates between the two actions and never achieves the desired sequence using UCB.

With our approach to deep exploration, the agent maintains an approximate posterior distribution of the optimal value functions $Q$. Initially, the agent's prior belief assigns a $1/2$ probability to either $Q(h, a_1) = 1$ or $Q(h, a_2) = 1$ for any $h$ that consistently applies $a_1$ or $a_2$, respectively, and assigns a value of 0 to all other history-action pairs. During any user life-cycle prior to receiving positive feedback, there is a $1/2$ probability that the agent will consistently apply action $a_2$ for $T$ times, thereby eliminating the possibility of $Q(h, a_1) = 1$. In expectation, the agent needs $2$ lifecycles to determine $Q(h, a_2) = 1$ when persistently applying $a_2$. Consequently, the agent requires only $\Theta(1)$ user life-cycle to learn the optimal policy using the deep exploration strategy.

\subsection{SeqRec Sample Complexity: Multi-User Case}
\textit{Building on Example 1, we now consider a SeqRec environment with $N$ users, denoted as $\mathcal{U} = \{1, 2, \dots, N\}$. For some users in $\mathcal{U}$, action $a_1$ increases satisfaction by 1, while for others, action $a_2$ does the same. For each user $u \in \mathcal{U}$, the action that improves satisfaction is deterministic, and we refer to this action as user $u$'s preferred action. We have $b_u = \tau_u = T$ for all $u \in \mathcal{U}$, and $N \ll 2^T$. The agent aims to optimize cumulative rewards across all users, with prior knowledge that either action $a_1$ or $a_2$ enhances each user's satisfaction level. We assume the existence of a function approximator capable of accurately modeling each user's preferred action after observing the preferred actions of $N/2$ users.} 

In this example, we extend Example 1 from a single-user environment to a multi-user setting, where learnings from one user can be generalized to serve other users as well. Our aim is to demonstrate that deep exploration remains crucial to an agent's success in a RS environment with a large user base, even when equipped with powerful generalization capabilities. Since the myopic UCB strategy is unable to identify the optimal policy even in single-user SeqRec, we exclude it from this analysis.

First, we consider the $\epsilon$-greedy and myopic Thompson sampling strategies. As an agent equipped with such strategies have a $\Theta(1/2^T)$ probability of learning the preferred action for any user $u$ in a life-cycle, the expected number of preferred actions learned after all $N$ users' first life-cycles is $\Theta(N / 2^T)$. In expectation, $\Theta(2^{T-1})$ life-cycles from all $N$ users are needed for the agent to learn all preferred actions. If the agent undergoes $M$ life-cycles and $M \leq 2^{T-1}$, the expected cumulative reward across the $N$ users in $\mathcal{U}$ is $\Theta(MN / 2^T)$. If $M > 2^{T-1}$, the expected cumulative reward is $\Theta(N(M - 2^{T-1}))$. 

Now, let's consider deep exploration. This algorithm has a $1/2$ probability of learning each user's preferred action after a single life-cycle of interaction. In expectation, after the first life-cycles of users in $\mathcal{U}$, the system can perfectly model all users' preferred actions, yielding an expected cumulative reward of $\Theta(NM)$ in $M$ life-cycles.

The analysis above highlights that intelligent exploration is not only crucial for learning single-user preferences but also for inferring population preferences. Exploration strategies, such as $\epsilon$-greedy, myopic Thompson sampling, and myopic UCB, are ill-suited for problems characterized by sparse and delayed consequences, even when paired with function approximators possessing strong generalization capabilities like deep neural networks. In contrast, deep exploration is much more adept at handling these types of problems.

In the following two sections, we will introduce neural network architectures that are useful for exploration and epistemic uncertainty estimation in RS.
\section{Epistemic Uncertainty Estimation via Epistemic Neural Networks (ENN)} \label{enn}
Deep neural networks have come a long way in enhancing marginal predictions. Over the past decade, significant advancements have been made in various architectures for RS, such as Neural Collaborative Filtering \cite{he2017neural}, Wide \& Deep \cite{cheng2016wide}, and Multi-Head Self-Attention \cite{wu2019neural}. These models provide a robust baseline for sophisticated representations of users, recommendations, and their interactions. However, these neural networks are unable to identify what they do not know, i.e., assessing the epistemic uncertainty concerning the environment or the user's preferences. In this section, we will focus on two neural network strategies for estimating epistemic uncertainty that are compatible with the concept of deep exploration.

\subsection{Deep Ensemble \cite{lu2017ensemble, osband2018randomized, osband2016deep}}
Deep ensemble is a neural network architecture that employs an ensemble of neural networks to estimate the posterior distribution of a target variable. This architecture is initialized with $M$ base neural networks, denoted as $\left\{f_{\beta_1}, \dots, f_{\beta_M}\right\}$, with each network parameterized by $\beta_1, \ldots, \beta_M$, respectively. Additionally, $M$ prior neural networks are included, represented as $\left\{f_{\beta_1^p}, \dots, f_{\beta^p_M}\right\}$.

The prior neural networks are initialized using Glorot sampling and remain unaltered throughout the model's lifespan \cite{osband2018randomized, glorot2010understanding}. These prior neural networks serve to regularize the posterior parameters, preventing the posterior distribution from collapsing too easily \cite{zhang2017understanding, bartlett2017spectrally}.

For a given input $x \in \mathbb{R}^{d_x}$, a deep ensemble generates a posterior sample estimate by sampling $z \sim \text{Unif}(\{1, \ldots, M\})$ and computing $\hat{y} = f_{\beta_z}(x) + \alpha f_{\beta^p_z}(x)$. Here, $f: \mathbb{R}^{d_x} \rightarrow \mathbb{R}$, and $\alpha \in (0, 1)$ is a scaling factor. This method is known as ensemble sampling \cite{lu2017ensemble}, which approximates Thompson sampling \cite{QinEnsembleSampling}.

In the context of supervised learning, given a loss function $\mathcal{L}$ and a target variable $y$, an ensemble optimizes each particle $\beta_z$ in the ensemble through
\begin{equation}\label{eq:ensemble}
    \beta_z = \argmin_{\beta} \mathcal{L}(y, f_{\beta}(x) + \alpha f_{\beta^p_z}(x)).
\end{equation}

For simplicity in notation, we denote $h_{\theta, \theta^p}(x, z) = f_{\beta_z}(x) + \alpha f_{\beta^p_z}(x)$ as the $z$th neural network in the ensemble, where $h: \mathbb{R}^{d_x} \times \{1, \dots, M\} \rightarrow \mathbb{R}$ and $\theta = (\beta_1, \dots, \beta_M), \theta^p = (\beta^p_1, \dots, \beta^p_M)$. 

\subsection{EpiNet \cite{osband2021epistemic}}

Although deep ensemble provides a valuable tool for epistemic uncertainty estimation, its scalability with large deep neural networks remains a challenge. There are two primary bottlenecks of deep ensembles. First, the number of parameters required is $M$ times larger than that of a single neural network. Second, a deep ensemble can only offer $M$ particles sampled from the approximate posterior distribution. EpiNet \cite{osband2021epistemic} proposes a potential solution to both of these bottlenecks.

EpiNet is an add-on architecture designed for general neural network architectures, enabling epistemic uncertainty estimation and improved joint predictions. Given a neural network function approximator $f_\beta: \mathbb{R}^{d_x} \rightarrow \mathbb{R}$, parameterized by $\beta$, EpiNet first extracts the last layer representation $\sigma_{\beta}(x)$, where $\sigma_\beta: \mathbb{R}^{d_x} \rightarrow \mathbb{R}^{d_\sigma}$. Next, EpiNet concatenates the representation with an epistemic index $z \in \mathbb{R}^{d_z}$ and executes a forward pass on the concatenated vector. The epistemic index can be a one-hot vector or sampled from a Gaussian prior. Similar to deep ensemble, EpiNet also maintains a main network $g_\eta$ and a prior network $g_{\eta^p}$, where $g: \mathbb{R}^{d_\sigma} \times \mathbb{R}^{d_z} \rightarrow \mathbb{R}^{d_z}$. Hence given a neural network $f_\beta$, an input feature vector $x$ and an epistemic index $z$, its target variable estimate $\hat{y}$ with the EpiNet add-on is 
\begin{equation}\label{eq:epinet}
    \hat{y} = f_\beta(x) + \left(g_\eta(sg[\sigma_\beta(x)], z) + \alpha g_{\eta^p}(sg[\sigma_\beta(x)], z)\right)^Tz.
\end{equation}
where $sg[\cdot]$ indicates stop-gradient in backpropagation. In a logistic regression case, $\hat{y}$ goes through an additional Sigmoid function. For simplicity of notation and consistency with the notations for deep ensemble, we denote Equation \ref{eq:epinet} by $h_{\theta, \theta^p}(x, z)$, where $\theta = (\beta, \eta), \theta^p = \eta^p$. During optimization, given a target variable $y$
\begin{equation}
    \theta = \argmin_{\theta} \sum_{z\in \mathcal{Z}}\mathcal{L}\left(y, h_{\theta, \theta^p}(x, z)\right).
\end{equation}
Parameters within $sg[\cdot]$ does not receive gradient flow. Here $\mathcal{Z}$ is a set of epistemic indices sampled from $z$'s prior to approximate the loss across the prior distribution of $z$. 

\section{A Deep Exploration Agent} \label{sec:algo}
In this section, we present an efficient implementation of deep exploration utilizing the Randomized Value Function (RVF) approach \cite{osband2019deep} through ENNs and Deep Q-learning. RVF exhibits two main features. First, it introduces perturbations to the received rewards, encouraging exploratory behavior. Another advantage of RVF is that, given a value function sampled from an ENN, an agent employing the RVF strategy consistently adheres to the policy induced by the value function for the entire user lifecycle, facilitating deep exploration.

\subsection{Representations} 
In this work, we assume that the agent has access to representations of each user, each recommendation, and the historical interactions between users and the RS. We denote the feature representation of a user $u$ as $\psi_u \in \mathbb{R}^{d_u}$ and that of an action $a$ as $\phi_a \in \mathbb{R}^{d_a}$. Both $\psi_u$ and $\phi_a$ are deterministic and remain constant over time. These representations can generally include, for example, demographic information about a user or a text embedding of a recommendation. Additionally, the agent is offered an interaction feature extractor $\textbf{extract\_interact\_features}: \mathcal{H} \times \mathcal{U} \rightarrow \mathbb{R}^{d_\xi}$ by the environment. We denote $\xi_{t, u} = \textbf{extract\_interact\_features}(h_t, u)$ to represent the historical interaction between user $u$ and the agent up to timestep $t$.

\subsection{Deep Q-learning}
Deep Q-learning has achieved remarkable successes in games \cite{mnih2015human, silver2017mastering} and control tasks \cite{schrittwieser2020mastering}, where it demonstrates super-human performance. In the context of RS, a deep Q-learning agent uses a neural network to estimate its value function, $Q_\beta: \mathbb{R}^{d_u} \times \mathbb{R}^{d_a} \times \mathbb{R}^{d_\xi} \rightarrow \mathbb{R}$, parameterized by $\beta$. Following the problem definition in Section \ref{prob}, at time step $t$, a deep Q-learning agent selects an action:
\begin{equation}
A_t = \left\{\left(u, \argmax_{a\in \mathcal{A}_{t, u}} Q_\beta(\psi_u, \phi_a, \xi_{t, u})\right) : u \in \mathcal{U}_t\right\},
\end{equation}
where $\mathcal{A}_{t, u} = c(O_{t, u})$. The agent explores by either modifying $Q_\beta$ or adopting a random exploration approach. For instance, when using $\epsilon$-Greedy as the exploration strategy, with probability $\epsilon$, the agent chooses a random action and follows the value function greedily otherwise. After executing action $A_t$, the agent receives a new observation $O_{t+1}$ from the environment and computes reward $R_{t+1}$. If $u \in \mathcal{U}_t \cap \mathcal{U}_{t+1}$, the agent stores a transition $(\psi_u, \phi_{A_{t,u}}, \xi_{t, u}, R_{t+1, u}$, $\xi_{t+1, u}, \mathcal{A}_{t+1, u})$, where $\mathcal{A}_{t+1, u} = c(O_{t+1, u})$, in its replay buffer $\mathcal{D}$ for future learning. It is worth noting that this approach can be easily extended to accommodate multi-step transitions by referencing the last observation generated by user $u$.

After several steps of interaction with the environment, the agent samples data from its replay buffer to improve its value function model $Q_\beta$. To stabilize temporal difference learning with deep neural networks, a deep Q-learning agent employs a target neural network strategy, creating a copy of the value function model as the target model $Q_{\beta'}$ and copying the parameters from $Q_\beta$ to $Q_{\beta'}$ every $K$ steps. More formally, with learning rate $\alpha$, the update for $\beta$ is:

\begin{equation}
\begin{split}
\beta \leftarrow &\beta - \alpha \nabla_{\beta} \sum_{u \in \mathcal{U}_t \cap \mathcal{U}_{t+1}}\Big[\Big(R_{t+1, u} + \\
&\max_{a \in \mathcal{A}_{t+1, u}} Q_{\beta'}(\psi_u, \phi_a, \xi_{t+1, u})
- Q_\beta(\psi_u, \phi_{A_{t, u}}, \xi_{t, u})\Big)^2\Big]\mid_{\beta = \beta_n}.\\
\end{split}
\end{equation}

It is important to note that we did not introduce the discount factor, which is common in RL literature, as we are optimizing for a fixed-horizon cumulative reward.

\subsection{Deep Exploration via Randomized Value Function and Epistemic Neural Network}

An agent can conduct deep exploration by employing Deep Q-learning through RVF and an ENN, $h^Q_{\theta, \theta^p}$, in order to estimate the posterior distribution of value functions. $h^Q_{\theta, \theta^p}$ takes a user representation, an action representation, an interaction feature vector and an epistemic index as input and outputs a value function sample from an approximate posterior distribution. At the beginning of a user life-cycle, a deep exploration agent samples an epistemic index $z_u \in \mathbb{R}^{d_z}$ from the prior distribution $P_z$, and maintains this index throughout the life-cycle. At time step $t$, the deep exploration agent then proceeds to follow a greedy approach: 
\begin{equation}
    A_t = \left\{\left(u, \argmax_{a\in \mathcal{A}_{t, u}}h^Q_{\theta, \theta^p}(\psi_u, \phi_a, \xi_{t, u}, z_u)\right) : u \in \mathcal{U}_t\right\}.
\end{equation}
See Section \ref{enn} for two potential architectures for $h^Q_{\theta, \theta^p}$. After taking action $A_t$, the agent makes a new observation $O_{t+1}$ from the environment and computes reward $R_{t+1}$. 

Similar to a deep Q-learning agent, a deep exploration agent also maintains replay buffers. To incentivize exploration, we adopt the reward perturbation approach in \cite{osband2019deep} when updating the buffers. There is a notable distinction when using a deep ensemble compared to an EpiNet: an agent utilizing a deep ensemble must maintain a separate perturbed buffer for each particle neural network within the ensemble, whereas an agent employing an EpiNet requires only a single perturbed buffer. To elaborate, if $u \in \mathcal{U}_{t} \cap \mathcal{U}_{t+1}$, an agent leveraging deep ensemble stores a transition $(\psi_u, \phi_{A_{t,u}}, \xi_{t, u}, R_{t+1, u} + W^z_{t+1, u}, \xi_{t+1, u}, \mathcal{A}_{t+1, u})$ in the $z$th buffer for future learning. Here, $\mathcal{A}_{t+1, u} = c(O_{t+1, u})$, $W^z_{t+1, u} \sim \mathcal{N}(0, \sigma^2)$ and $\sigma^2$ is a hyperparameter. On the other hand, an agent employing an EpiNet only maintains a single replay buffer and stores a single transition with $W_{t+1, u} \sim \mathcal{N}(0, \sigma^2)$.

A deep exploration agent then samples from its buffer to update its parameter $\theta$. With learning rate $\alpha$, $\theta$ is updated via
\begingroup\makeatletter\def\f@size{8}\check@mathfonts
\def\maketag@@@#1{\hbox{\m@th\large\normalfont#1}}%
\begin{equation}\label{eq:update}
\begin{split}
    \theta \leftarrow &\theta - \alpha \nabla_{\theta} \sum_{z\in \mathcal{Z}}\sum_{u \in \mathcal{U}_t \cap \mathcal{U}_{t+1}}\Big[\Big(\tilde{R}_{t+1, u, z} + \\
    &\max_{a \in \mathcal{A}_{t+1, u}} h^Q_{\theta', \theta^p}(\psi_u, \phi_a, \xi_{t+1, u}, z)
     - h^Q_{\theta, \theta^p}(\psi_u, \phi_{A_{t,u}}, \xi_{t, u}, z)\Big)^2\Big]\mid_{\theta = \theta_n}.\\
\end{split}
\end{equation}
\endgroup
Here, $\tilde{R}_{t+1, u, z}$ denotes the perturbed reward and $h^Q_{\theta', \theta^p}$ represents the target network. For an agent utilizing EpiNet, $\mathcal{Z}$ is a set of epistemic indices sampled from $z$'s prior, which helps approximate the loss across the entire distribution of $z$. Moreover, $\tilde{R}_{t+1, u, z}$ remains consistent across all $z$ values. Conversely, for an agent that employs a deep ensemble, $\tilde{R}_{t+1, u, z}$ corresponds to the perturbed reward from the $z$th replay buffer in the ensemble, and $\mathcal{Z}$ is defined as $\{1, \ldots, M\}$.

For a comprehensive view of the deep exploration algorithm with EpiNet, please refer to Algorithm \ref{alg:de}.
and Figure \ref{fig:deep_explore} for visualization. 
\begingroup\makeatletter\def\f@size{8}\check@mathfonts
\def\maketag@@@#1{\hbox{\m@th\large\normalfont#1}}%
\begin{figure}[t]
     \centering
     \includegraphics[width=0.5\textwidth]{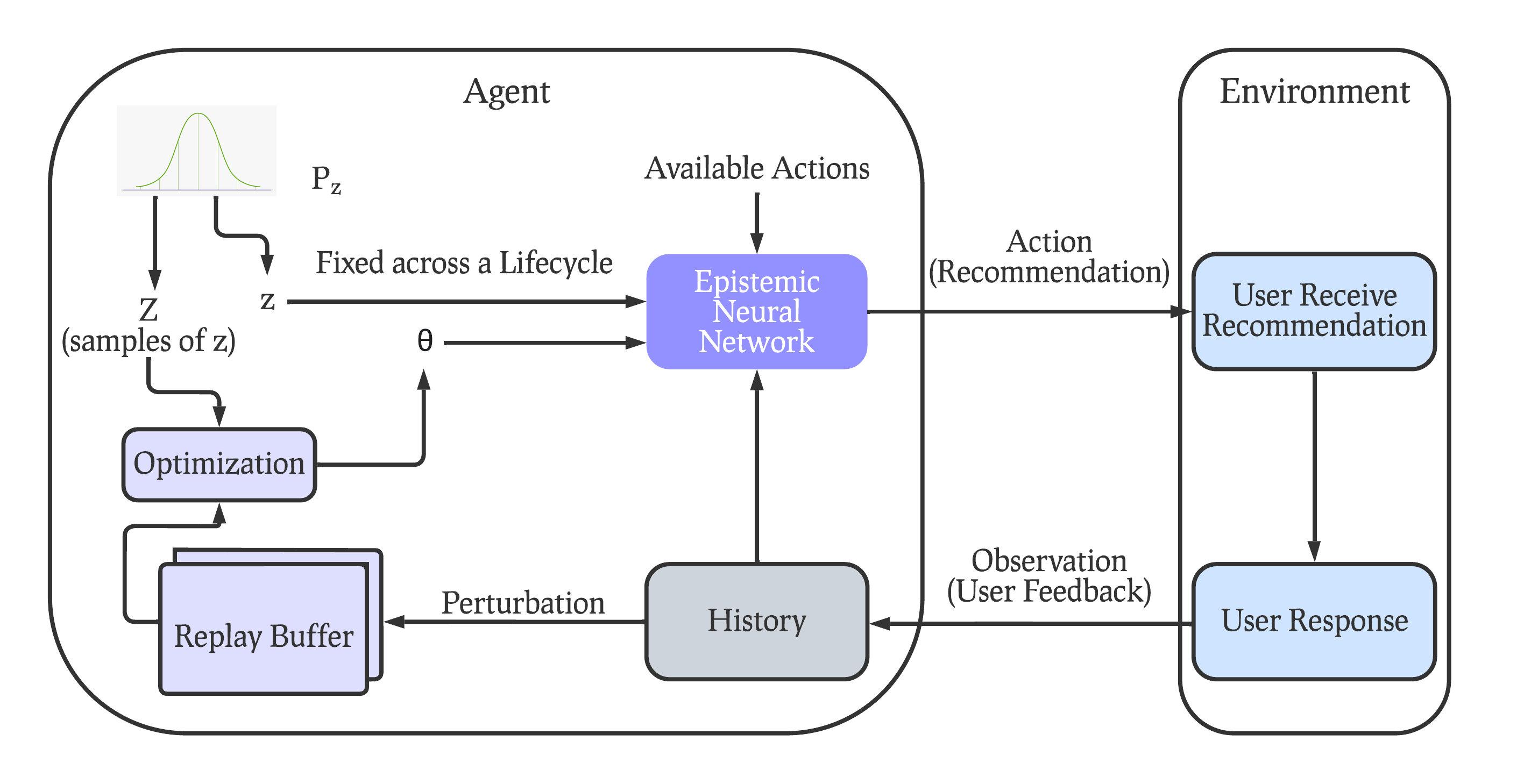}
     \caption{Deep Exploration Agent Architecture for Recommendation Systems}
     \label{fig:deep_explore}
\end{figure}
\endgroup

\begin{figure*}[t]
\begin{minipage}{1\textwidth}
\begin{algorithm}[H]
\small
	\caption{Deep Exploration for RS with EpiNet}\label{alg:de}
	\begin{algorithmic}[1]
	    \State Initialize epistemic neural network $h^Q_{\theta, \theta^p}$ and its target $h^Q_{\theta', \theta^p}$ via EpiNet and initialize reward variance $\sigma^2$.
		\State Initialize replay buffer $\mathcal{D}$.
            \For{$t = 1, 2, \dots$} \Comment{Progressing over time steps}
                \State Makes observation $O_{t,u}$ and computes $R_{t,u}, \forall u \in \mathcal{U}_t$.
                \For{$u \in \mathcal{U}_t$}
                    \If{$z_u$ is not set $\lvert\rvert$ $\mathcal{A}_{t,u} = \{\text{no-op}\}$}
                        \State $z_u \sim P_z$
                    \EndIf
                    \State $\xi_{t, u} = \textbf{extract\_interact\_features}(H_t, u)$ 
                    \State $\mathcal{A}_{t,u} = c(O_{t,u})$.
                    \State $A_{t,u} = \arg\max_{a \in \mathcal{A}_{t, u}}h^Q_{\theta, \theta^p}(\psi_u, \phi_a, \xi_{t, u}, z_u)$.
                    \If{$u \in \mathcal{U}_{t-1}$}
                        \State Sample $W_{t,u} \sim \mathcal{N}(0, \sigma^2)$.
                        \State Store $(\psi_u, \phi_{A_{t-1, u}}, \xi_{t-1, u}, R_{t, u} + W_{t, u}, \xi_{t, u}, \mathcal{A}_{t, u})$ in $\mathcal{D}$.
                    \EndIf
                \EndFor
                \State Sample data from $\mathcal{D}$ and use Equation \ref{eq:update} to update $\theta$.
            \EndFor
    \end{algorithmic}
\end{algorithm}
\vspace{-0.1in}
\end{minipage}
\hfill
\begin{minipage}{1\textwidth}
\vspace{-0.1in}
\end{minipage}
\end{figure*}


\section{Experiments}

\begingroup\makeatletter\def\f@size{8}\check@mathfonts
\def\maketag@@@#1{\hbox{\m@th\large\normalfont#1}}%
\begin{table*}[ht!]
\small
  \caption{Average Life-Cycle Cumulative Reward across Users (DE Stands for Deep Exploration)}
  \label{tab:exp_reward}
  \resizebox{\textwidth}{!}{
  \begin{tabular}{cccccc}
    \toprule
    Algorithm & Toy & E-Commerce Train & E-Commerce Eval & Multi-User Slate Ranking Train & Multi-User Slate Ranking Eval\\
    \midrule 
    $\epsilon$-Greedy & $0.0 \pm 0.0$ & $1.07 \pm 0.02$ & $0.95 \pm 0.07$ & $94.79 \pm 1.21$\ & $112.24 \pm 2.87$\\
    Neural TS & $0.0 \pm 0.0$ & $6.55 \pm 1.14$ & $6.37 \pm 0.90$ & $111.323 \pm 1.34$ & $119.60 \pm 5.22$\\
    Neural UCB & $0.0 \pm 0.0$ & $9.13 \pm 0.04$ & $8.035 \pm 0.15$ & $117.341 \pm 2.03$ & $119.14 \pm 3.70$\\
    Neural LinUCB & $0.0 \pm 0.0$ & $9.42 \pm 0.13$ & $8.78 \pm 0.17$ & $121.344 \pm 4.40$ & $125.50 \pm 6.67$\\
    \midrule
    Ensemble - DE & $0.59 \pm 0.03$ & $11.46 \pm 0.83$ & $9.82 \pm 0.50$ & $135.844 \pm 3.62$ & $152.25 \pm 10.17$\\
    EpiNet - DE & $\mathbf{0.71 \pm 0.15}$ & $\mathbf{16.64 \pm 1.04}$ & $\mathbf{13.83 \pm 0.25}$ & $\mathbf{147.303 \pm 7.11}$ & $\mathbf{178.92 \pm 20.61}$\\
  \bottomrule
\end{tabular}}
\end{table*}
\endgroup

In this section, we conduct a series of experiments to demonstrate the benefits of deep exploration. We present a toy experiment following the setup in Example \ref{exmp:case_study}, a large-scale experiment based on an e-commerce environment, and another extensive experiment for slate ranking with parallel user streams, grounded in a game recommender system. These large-scale environments, designed by industry leaders, aim to ensure high-fidelity, model-based evaluation grounded on real-world data.

We compare deep exploration agents based on both deep ensemble and EpiNet against Q-learning agents that utilize strategies such as $\epsilon$-greedy, Neural Thompson Sampling (Neural TS) \cite{zhang2020neural}, Neural UCB \cite{zhou2019neural}, and Neural LinUCB \cite{xu2021neural}.

In all experiments, the scaling factor of the prior network is set to 0.3. When using ensemble sampling, we set the ensemble's cardinality to 10. In the case of EpiNet, the number of epistemic indices during optimization is set to 50, the dimension of the epistemic index is set to 10, and these indices are drawn from a standard Gaussian distribution.
All agents share the same value function architecture, featuring a single hidden layer of 20 units for the toy experiment, and hidden layers of size (200, 100, 50, 25) for the remaining experiments. All experiments are executed on AWS p4d24.xlarge instances equipped with two A100 GPUs.

The results of these experiments are presented in Table \ref{tab:exp_reward}.

\subsection{SeqRec Experiment}
We commence with an experiment utilizing the SeqRec environment introduced in Example \ref{exmp:case_study}. We've set $\mathcal{U}_t = {1}, \forall t$, with $\tau_1 = 10$, $b_1 = 10$, $g(1, a_1) = 0$, and $g(1, a_2) = 1$. Contrary to the case study Example 1 in Section \ref{example-1}, we do not implicitly inject prior knowledge into the agent. The action features are provided to the agent as one-hot feature vectors and each entry of the interaction feature $\xi_{t, u} \in \mathbb{R}^{10}$ is defined as: 
\begin{equation}
    \xi_{t, u}[i] = 
        \begin{cases} 
            -1 &\mbox{if } i > L_{t, u} \\ 
            1 & \mbox{if $A_{t-L_t+i, u} = a_1$ at time step $t - L_t + i$} \\
            0 & \mbox{otherwise} \end{cases},
\end{equation}
where $L_{t, u}$ is the user's current life-cycle length. Since there is a single user in this environment, the agent is not offered user representations. The results of the experiment, averaged over ten random seeds each with 100 life-cycles, are presented in the first column of Table \ref{tab:exp_reward}. The results indicate that only the deep exploration agent was able to learn a reasonable policy. 

\begin{figure*}
    \centering
     \begin{subfigure}[t]{0.3\textwidth}
         \includegraphics[width=\textwidth, valign=t]{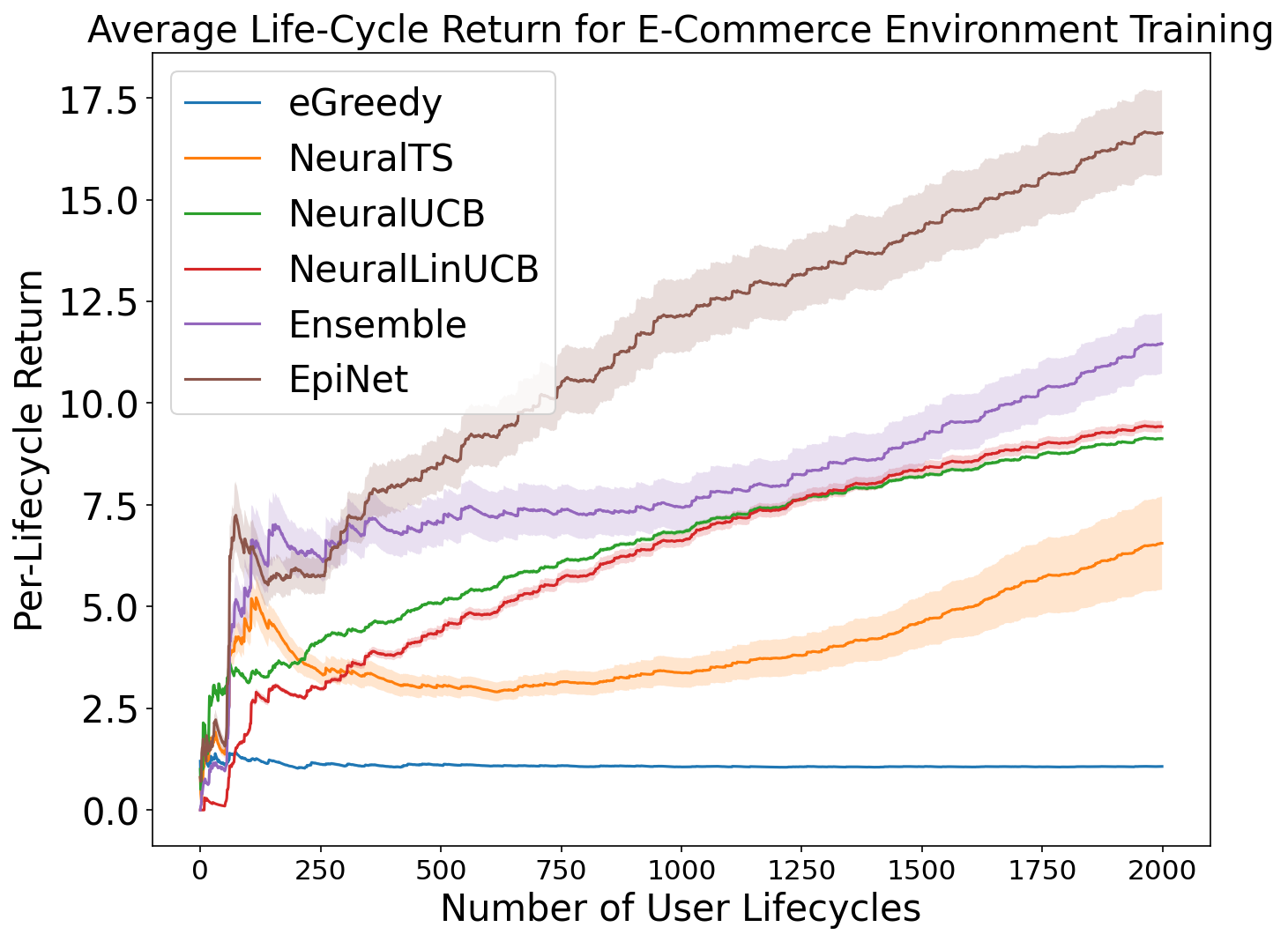}
         \caption{Average Life-Cycle Return for High-Fidelity E-Commerce Environment through Virtual-Taobao, 200 Users Each Serving 10 Times Sequentially}
         \label{fig:e_commerce_small}
     \end{subfigure}
     \hfill
     \begin{subfigure}[t]{0.3\textwidth}
         \includegraphics[width=\textwidth, valign=t]{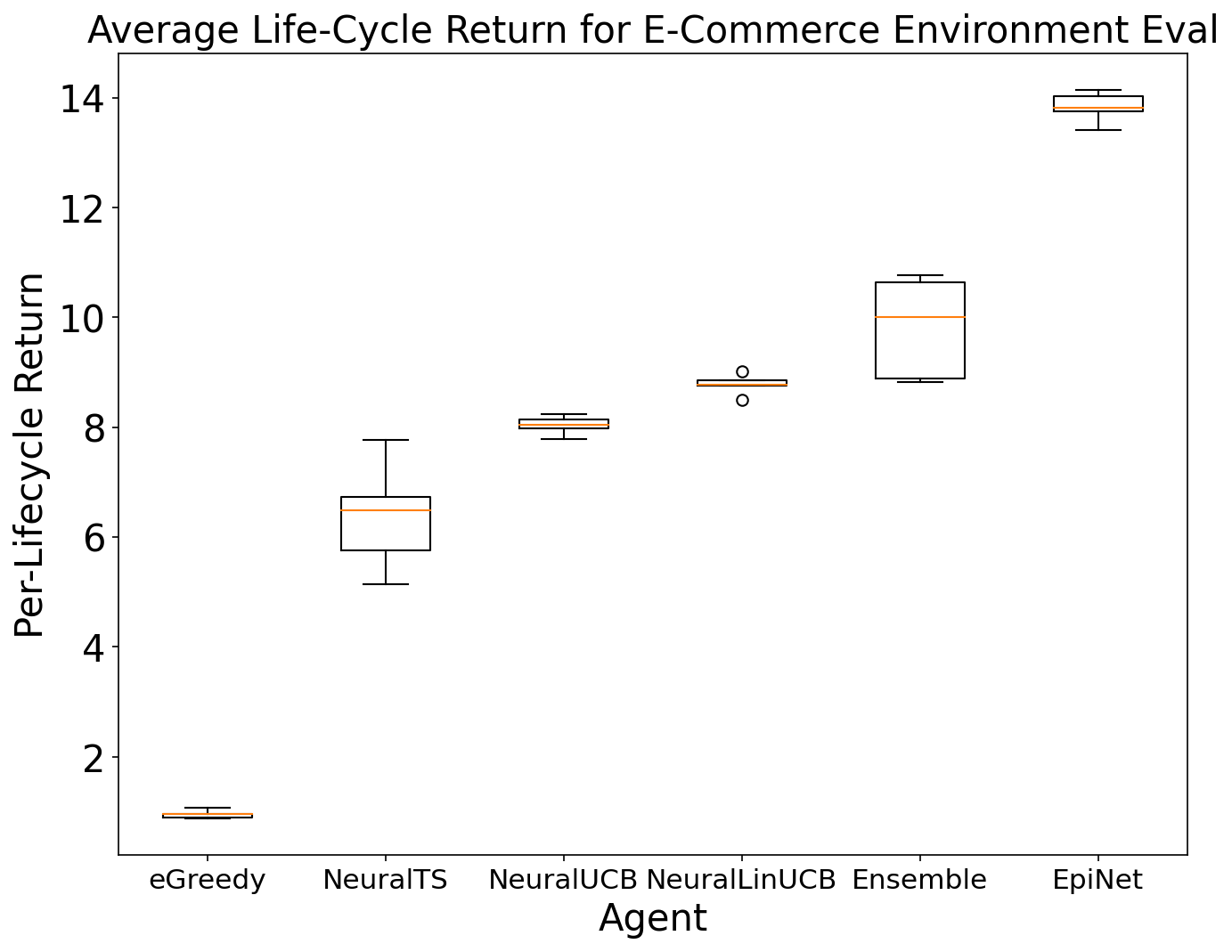}
         \caption{Average Life-Cycle Return for High-Fidelity E-Commerce Environment with 200 Cold Start Users that Agent Has Never Seen Before}
         \label{fig:e_commerce_eval}
     \end{subfigure}
     \hfill
     \begin{subfigure}[t]{0.3\textwidth}
         \includegraphics[width=\textwidth, valign=t]{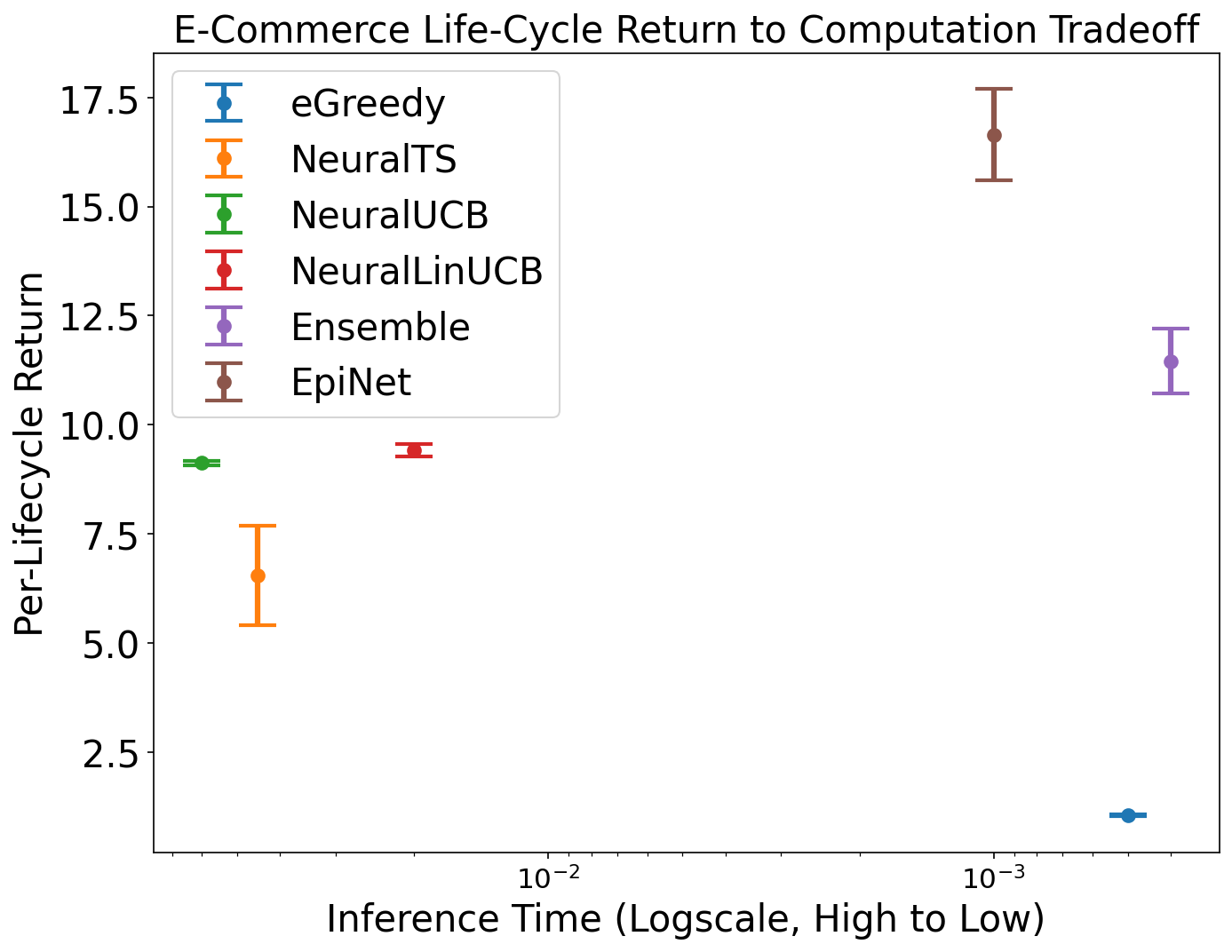}
         \caption{Tradeoff between Life-Cycle Return and Computation for E-Commerce Experiment. Better Tradeoff towards Top Right.}
         \label{fig:e_commerce_perf}
     \end{subfigure}
    \hfill
     \begin{subfigure}[t]{0.3\textwidth}
         \includegraphics[width=\textwidth, valign=t]{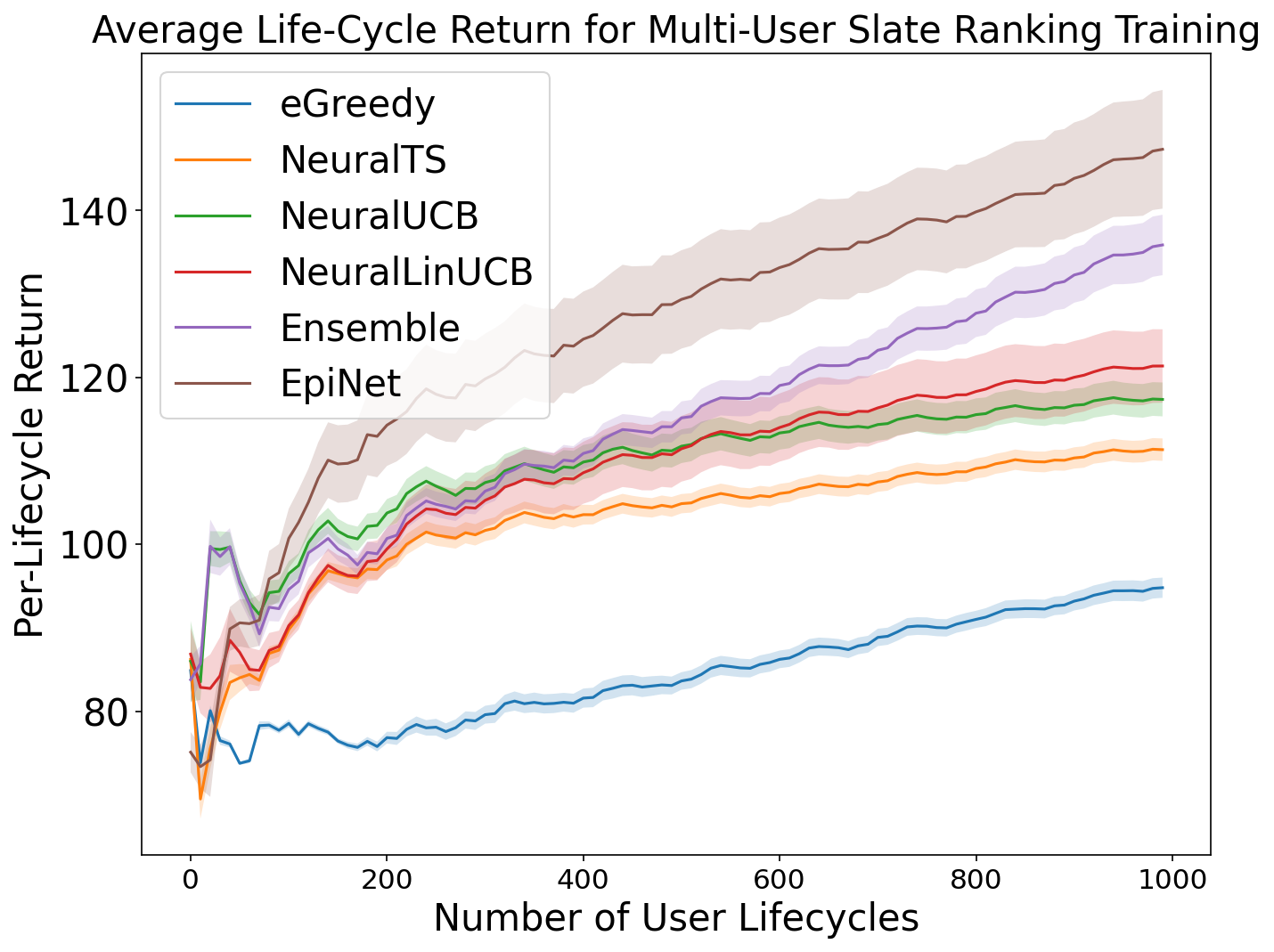}
         \caption{Average Life-Cycle Return for High-Fidelity Multi-User Slate Ranking Environment, 100 Users Each Served 10 Times with 10-User Parallel Streams}
         \label{fig:slate_small}
     \end{subfigure}
     \hfill
     \begin{subfigure}[t]{0.3\textwidth}
         \includegraphics[width=\textwidth, valign=t]{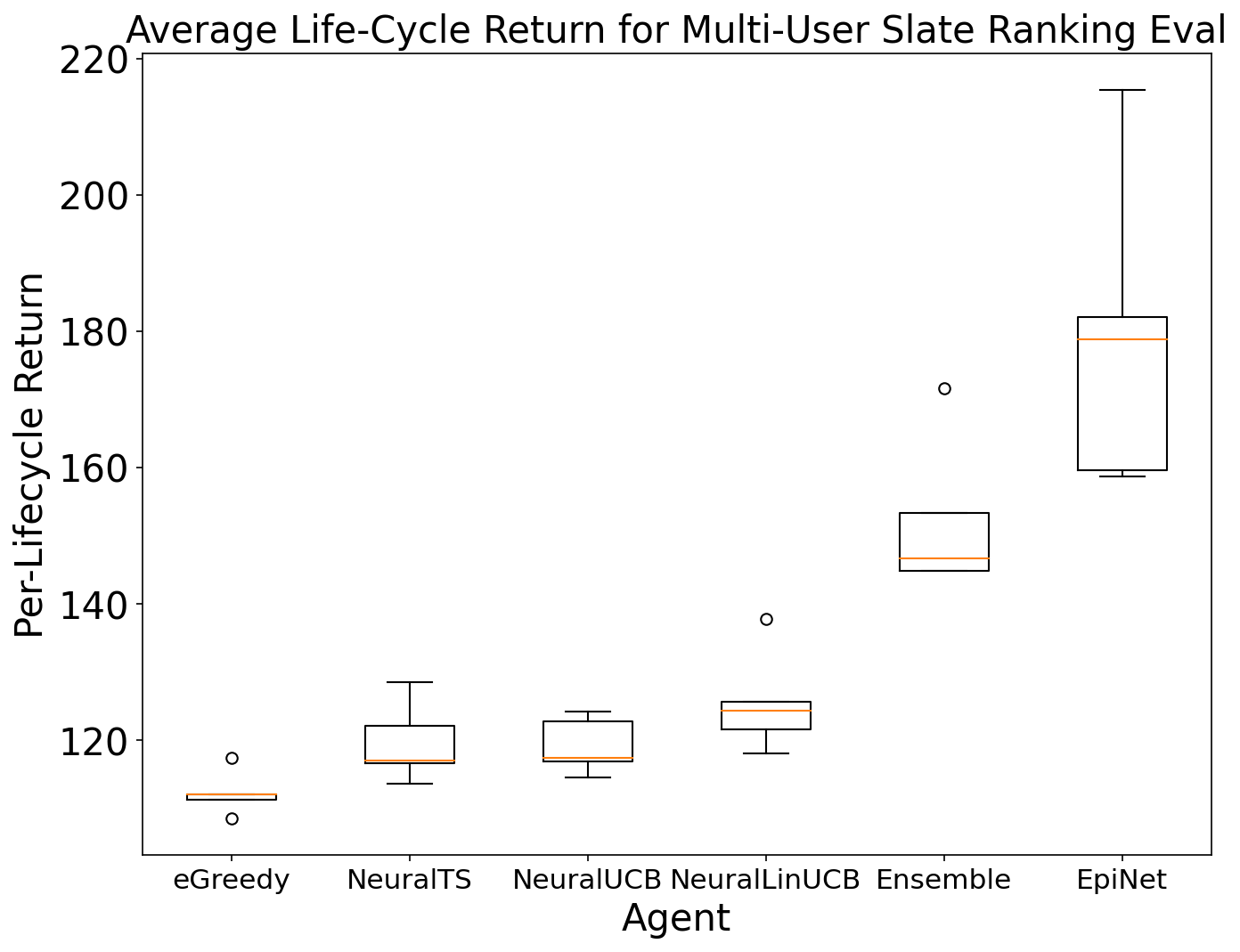}
         \caption{Average Life-Cycle Return for High-Fidelity Multi-User Slate Ranking Environment with 100 Cold Start Users that Agent Has Never Seen Before in 10-User Parallel Streams}
         \label{fig:slate_eval}
     \end{subfigure}
     \hfill
     \hspace{8pt}
     \begin{subfigure}[t]{0.3\textwidth}
         \includegraphics[width=\textwidth, valign=t]{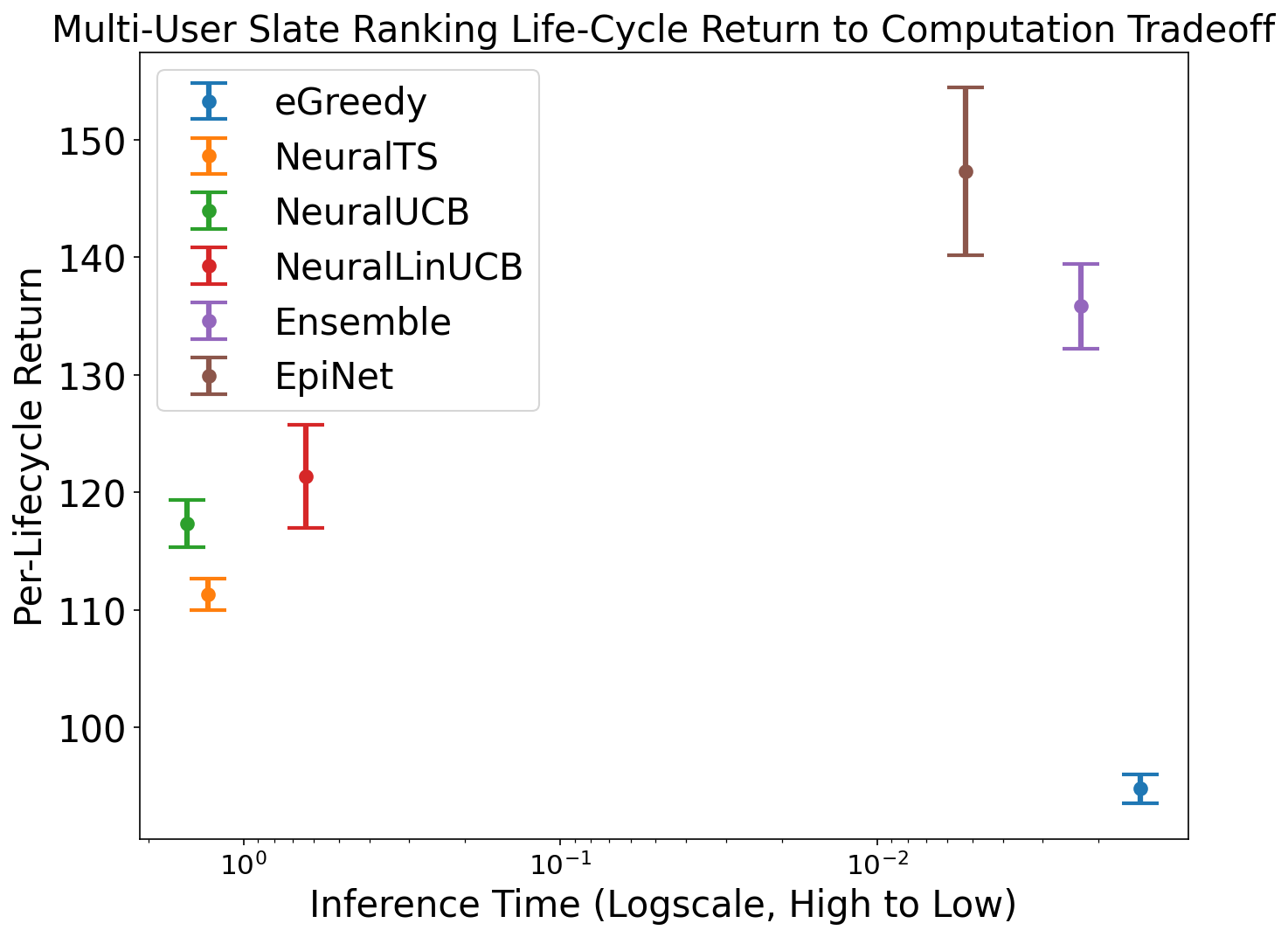}
         \caption{Tradeoff between Life-Cycle Return and Computation for Multi-User Slate Ranking Experiment. Better Tradeoff towards Top Right.}
         \label{fig:slate_perf}
     \end{subfigure}
     \caption{Experiment Results across E-Commerce and Multi-User Slate Ranking Environments. Average life-cycle return in each of the figures above plots $\frac{1}{\lvert \mathcal{U} \rvert}\sum_{u\in \mathcal{U}}\sum_{t=1}^T R_{t, u}$ averaged over 30 distinct seeds, where $\mathcal{U} = \bigcup^{T}_{t=1} \mathcal{U}_t$ is the community of all users the agent has served and $R_{t, u}$ is the reward at time step $t$ for user $u$. Note that since the Eval set is a completely new set of users, it is normal to see evaluation's average life-cycle return higher than training.}
\end{figure*}


\subsection{E-Commerce Environment} \label{sec:ecomm}
The E-Commerce environment, a classic context for Recommender Systems (RS), involves a platform suggesting items for user purchase, followed by user feedback via clicks or purchases. Virtual-Taobao \cite{shi2019virtual}, a model-based simulator of user behavior on one of the world's largest E-Commerce platforms, allows us to evaluate agent performance in an online scenario with high fidelity. The environment delivers user feedback and termination via Multi-Agent Adversarial Imitation Learning (MAIL) \cite{song2018multi}.

At the onset of each episode, the simulator samples a user from the user distribution via a Generative Adversarial Network (GAN) \cite{goodfellow2020generative}, subsequently providing the agent with the user's representation vector. We have arranged the environment so that at each time step, the agent is presented with 100 recommendation candidates, each with its representative features. The agent's task is to select one to propose to the user. Additionally, the agent has access to an interaction feature vector and each user's representative feature vector.

It is important to note that, due to the computational limitations of neural Thompson sampling and neural UCB—both of which necessitate inverting covariance matrices with the width of the neural network's full parameter size—we only utilize the last layer's parameter matrix and gradients for these methods in this experiment. The results of the subsequent experiments are detailed in Table \ref{tab:exp_reward}.

Refer to Figure \ref{fig:e_commerce_small}, which displays the results of an experiment involving 30 seeds serving 200 users, each engaged 10 life-cycles with the agent with each life-cycle happening sequentially. It is clear that both deep exploration strategies surpass other exploration methods. Notably, the deep exploration methods achieve reasonable performance much more rapidly than myopic exploration methods. Comparing the two deep exploration strategies, it is evident that EpiNet outperforms Ensemble Sampling, continually enhancing its performance throughout its interactions with users.

In addition to in-sample recommendations, we also evaluate the performance of agents trained online with 200 out-of-sample users, without any additional training. Please refer to Figure \ref{fig:e_commerce_eval}. The deep exploration methods, particularly the EpiNet-based deep exploration, significantly outperform all other candidates in out-of-sample evaluations.

Finally, the computation-performance trade-off of the algorithms is illustrated in Figure \ref{fig:e_commerce_perf}. In this graph, candidates situated towards the top right corner exhibit the optimal trade-off criteria. Here, again, the deep exploration strategies demonstrate a clear advantage.

\subsection{Slate Ranking with Parallel User Streams}
In an effort to extend our experimentation to a more complex and realistic setting, we utilize RL4RS \cite{wang2021rl4rs}, a real-world environment that provides parallel user streaming and slate ranking. User behavior predictions are generated via the Deep Interest Evolution Network (DIEN) \cite{zhou2019deep}, and users are sampled from a real dataset with anonymized and abstract feature representations.

In this environment, the agent interacts with 10 users simultaneously, providing each user with a separate slate of recommendations. Whenever a user concludes their interactions with the agent, the environment samples a new user to maintain a constant stream of 10 active users. At each time step, the agent is presented with 284 different combinations (slates) of recommendation items as its available actions. Each slate is represented by a 128-dimensional feature vector.

Similar to the E-Commerce experiment, for the sake of computational tractability, we resort to the inversion of the last layer gradients and parameter matrix for NeuralTS and NeuralUCB. The results of the subsequent experiments are documented in Table \ref{tab:exp_reward}.

Refer to Figure \ref{fig:slate_small} for a 30-seed experiment serving 100 users, each with 10 life-cycles. The results illustrate that both deep exploration strategies outperform other exploration methods in this scenario. When comparing the deep exploration strategies, we observe that EpiNet slightly surpasses ensemble sampling in performance.

Similar to the e-commerce experiment previously discussed, we also evaluate our agents on an additional set of 100 out-of-sample users, as shown in Figure \ref{fig:e_commerce_eval}. Given that the experiment results originate from a distinct set of users, it is expected that the average per-user return for the evaluation set may be higher than during training. From this figure, we observe that both deep exploration methods significantly outperform the competition when applied to out-of-sample users.

Lastly, we provide a visualization of the computation-performance tradeoff in Figure \ref{fig:e_commerce_perf}. Here, it becomes evident that both deep exploration strategies lead the field in terms of this tradeoff.

\section{Discussion}
Deep exploration's impact extends beyond better optimization of long-term cumulative reward. It has significant implications on data efficiency, eliminating model bias as well as data privacy. 
\subsection{Data Efficiency.}
Reinforcement learning has a notorious sample complexity \cite{kakade2003sample}, and sparse rewards only exacerbate this issue. As the internet and metaverse continue to expand, personalization becomes more expensive as more data must be stored for each use case. As the industry's biggest corporations and content providers, such as Meta and Google, face shortages in data centers and computing power, data efficiency becomes crucial for RS services. Deep exploration enables content providers to identify users' preferences with a limited number of interactions, even when feedback is sparse and delayed.
\subsection{Fairness.}
Fairness in gender, race and many other demographics has been one of the most important problems in machine learning research \cite{chouldechova2018frontiers, corbett2018measure, barocas2017fairness}. Demographic groups with a lot of data collected allow a machine learning model to generate better predictions than those with sparse data. This also applies to the RS field. Minorities enjoy less personalization because less data is collected from the group. Such phenomenon is magnified as we extend supervised learning RS algorithms to reinforcement learning and it becomes even worse in environments with sparse and delays in feedback. Deep exploration breaks the vicious cycle where the minorities get less data collected, are less personalized, then use the product less and hence get even less data collected. The low sample complexity provided by deep exploration allows for fast personalization even in environments where little data or feedback is collected. 
\subsection{Data Privacy.}
The law pushes for a stronger regulation on data privacy and less retention of data owned by content providers. Under such regulations, content providers no longer own infinite amount of data from each user and need to make use of every piece of data possible. Such regulations have made significant impact improving the safety of user data, but meanwhile adversely impact the quality of personalization provided by content providers. Furthermore, the rise of on-device computing that further protects user privacy by owning user data locally, which further reduces the amount of data owned by content providers for personalization. Deep Exploration serves as a great tool to counter such dilemma, leveraging strategic actions to learn users' preferences in a small number of interactions while allowing the community to improve user privacy and safety.
\section{Conclusion}
While deep exploration algorithms have shown great improvement over other baselines in experiments and have relatively efficient computational complexity, there is still room for future work to advance these algorithms in scalability, rigorous information seeking, and productionalization. In our experiments, we observed that deep exploration strategies tend to be slightly more expensive in both training and inference than vanilla neural networks, despite being faster than other exploration strategies. Additionally, more advanced exploration technologies, such as Information Directed Sampling (IDS) \cite{russo2014learning}, could be better information-seeking strategies for deep exploration. However, IDS remains completely intractable for real-world environments as it consumes exponentially more computational resources. To develop better information-seeking deep exploration algorithms, the community needs to invest in scalable and rigorous algorithms, as well as neural network architectures that can scale such algorithms' policy updates.

In this paper, we addressed the issue of sparse and delayed feedback in RS and explored the potential for deep exploration to improve data efficiency in personalization. We formally defined the RS problem in the context of sequential decision-making and presented real-world experimental results with our deep exploration algorithms in different high-fidelity RS environments. We demonstrated promise, through these experiments, in improving data efficiency and amplifying the rate of positive feedback in a scalable manner relative to other exploration designs in RS. 
We hope that the results and discussion of this paper will inspire the engineering of production systems that leverage deep exploration and show impact on real-world platforms through data-efficient personalization.

\bibliographystyle{ACM-Reference-Format}
\bibliography{sn-bibliography}


\begin{thebibliography}{66}


\ifx \showCODEN    \undefined \def \showCODEN     #1{\unskip}     \fi
\ifx \showDOI      \undefined \def \showDOI       #1{#1}\fi
\ifx \showISBNx    \undefined \def \showISBNx     #1{\unskip}     \fi
\ifx \showISBNxiii \undefined \def \showISBNxiii  #1{\unskip}     \fi
\ifx \showISSN     \undefined \def \showISSN      #1{\unskip}     \fi
\ifx \showLCCN     \undefined \def \showLCCN      #1{\unskip}     \fi
\ifx \shownote     \undefined \def \shownote      #1{#1}          \fi
\ifx \showarticletitle \undefined \def \showarticletitle #1{#1}   \fi
\ifx \showURL      \undefined \def \showURL       {\relax}        \fi
\providecommand\bibfield[2]{#2}
\providecommand\bibinfo[2]{#2}
\providecommand\natexlab[1]{#1}
\providecommand\showeprint[2][]{arXiv:#2}

\bibitem[Auer et~al\mbox{.}(2002)]%
        {auer2002finite}
\bibfield{author}{\bibinfo{person}{Peter Auer}, \bibinfo{person}{Nicolo
  Cesa-Bianchi}, {and} \bibinfo{person}{Paul Fischer}.}
  \bibinfo{year}{2002}\natexlab{}.
\newblock \showarticletitle{{Finite-Time Analysis of the Multiarmed Bandit
  Problem}}.
\newblock \bibinfo{journal}{\emph{Machine Learning}} \bibinfo{volume}{47},
  \bibinfo{number}{2-3} (\bibinfo{year}{2002}), \bibinfo{pages}{235--256}.
\newblock


\bibitem[Barocas et~al\mbox{.}(2017)]%
        {barocas2017fairness}
\bibfield{author}{\bibinfo{person}{Solon Barocas}, \bibinfo{person}{Moritz
  Hardt}, {and} \bibinfo{person}{Arvind Narayanan}.}
  \bibinfo{year}{2017}\natexlab{}.
\newblock \showarticletitle{{Fairness in Machine Learning}}.
\newblock \bibinfo{journal}{\emph{NeurIPS tutorial}}  \bibinfo{volume}{1}
  (\bibinfo{year}{2017}), \bibinfo{pages}{2}.
\newblock


\bibitem[Bartlett et~al\mbox{.}(2017)]%
        {bartlett2017spectrally}
\bibfield{author}{\bibinfo{person}{Peter~L Bartlett}, \bibinfo{person}{Dylan~J
  Foster}, {and} \bibinfo{person}{Matus~J Telgarsky}.}
  \bibinfo{year}{2017}\natexlab{}.
\newblock \showarticletitle{{Spectrally-Normalized Margin Bounds for Neural
  Networks}}. In \bibinfo{booktitle}{\emph{Advances in Neural Information
  Processing Systems}}. \bibinfo{pages}{6240--6249}.
\newblock


\bibitem[Blanda(2016)]%
        {blanda2016online}
\bibfield{author}{\bibinfo{person}{Stephanie Blanda}.}
  \bibinfo{year}{2016}\natexlab{}.
\newblock \showarticletitle{Online {R}ecommender {S}ystems--{H}ow {D}oes a
  {W}ebsite {K}now {W}hat {I} {W}ant?}
\newblock \bibinfo{journal}{\emph{American Mathematical Society. Retrieved
  October}}  \bibinfo{volume}{31} (\bibinfo{year}{2016}).
\newblock


\bibitem[Bouneffouf et~al\mbox{.}(2012)]%
        {bouneffouf2012contextual}
\bibfield{author}{\bibinfo{person}{Djallel Bouneffouf}, \bibinfo{person}{Amel
  Bouzeghoub}, {and} \bibinfo{person}{Alda~Lopes Gan{\c{c}}arski}.}
  \bibinfo{year}{2012}\natexlab{}.
\newblock \showarticletitle{{A Contextual-Bandit Algorithm for Mobile
  Context-Aware Recommender System}}. In
  \bibinfo{booktitle}{\emph{International Conference on Neural Information
  Processing}}. Springer, \bibinfo{pages}{324--331}.
\newblock


\bibitem[Brod{\'e}n et~al\mbox{.}(2018)]%
        {broden2018ensemble}
\bibfield{author}{\bibinfo{person}{Bj{\"o}rn Brod{\'e}n},
  \bibinfo{person}{Mikael Hammar}, \bibinfo{person}{Bengt~J Nilsson}, {and}
  \bibinfo{person}{Dimitris Paraschakis}.} \bibinfo{year}{2018}\natexlab{}.
\newblock \showarticletitle{{Ensemble Recommendations via Thompson Sampling: an
  Experimental Study within E-Commerce}}. In \bibinfo{booktitle}{\emph{23rd
  international conference on intelligent user interfaces}}.
  \bibinfo{pages}{19--29}.
\newblock


\bibitem[Chapelle and Li(2011)]%
        {chapelle2011empirical}
\bibfield{author}{\bibinfo{person}{Olivier Chapelle} {and}
  \bibinfo{person}{Lihong Li}.} \bibinfo{year}{2011}\natexlab{}.
\newblock \showarticletitle{{An Empirical Evaluation of Thompson Sampling}}. In
  \bibinfo{booktitle}{\emph{Advances in Neural Information Processing
  Systems}}. \bibinfo{pages}{2249--2257}.
\newblock


\bibitem[Chen et~al\mbox{.}(2019a)]%
        {chen2019top}
\bibfield{author}{\bibinfo{person}{Minmin Chen}, \bibinfo{person}{Alex Beutel},
  \bibinfo{person}{Paul Covington}, \bibinfo{person}{Sagar Jain},
  \bibinfo{person}{Francois Belletti}, {and} \bibinfo{person}{Ed~H Chi}.}
  \bibinfo{year}{2019}\natexlab{a}.
\newblock \showarticletitle{Top-{K} {O}ff-policy {C}orrection for a {REINFORCE}
  {R}ecommender {S}ystem}. In \bibinfo{booktitle}{\emph{Proceedings of the
  Twelfth ACM International Conference on Web Search and Data Mining}}.
  \bibinfo{pages}{456--464}.
\newblock


\bibitem[Chen et~al\mbox{.}(2021)]%
        {chen2021values}
\bibfield{author}{\bibinfo{person}{Minmin Chen}, \bibinfo{person}{Yuyan Wang},
  \bibinfo{person}{Can Xu}, \bibinfo{person}{Ya Le}, \bibinfo{person}{Mohit
  Sharma}, \bibinfo{person}{Lee Richardson}, \bibinfo{person}{Su-Lin Wu}, {and}
  \bibinfo{person}{Ed Chi}.} \bibinfo{year}{2021}\natexlab{}.
\newblock \showarticletitle{{Values of User Exploration in Recommender
  Systems}}. In \bibinfo{booktitle}{\emph{Fifteenth ACM Conference on
  Recommender Systems}}. \bibinfo{pages}{85--95}.
\newblock


\bibitem[Chen et~al\mbox{.}(2022)]%
        {chen2022off}
\bibfield{author}{\bibinfo{person}{Minmin Chen}, \bibinfo{person}{Can Xu},
  \bibinfo{person}{Vince Gatto}, \bibinfo{person}{Devanshu Jain},
  \bibinfo{person}{Aviral Kumar}, {and} \bibinfo{person}{Ed Chi}.}
  \bibinfo{year}{2022}\natexlab{}.
\newblock \showarticletitle{{Off-Policy Actor-critic for Recommender Systems}}.
  In \bibinfo{booktitle}{\emph{Proceedings of the 16th ACM Conference on
  Recommender Systems}}. \bibinfo{pages}{338--349}.
\newblock


\bibitem[Chen et~al\mbox{.}(2019b)]%
        {chen2019generative}
\bibfield{author}{\bibinfo{person}{Xinshi Chen}, \bibinfo{person}{Shuang Li},
  \bibinfo{person}{Hui Li}, \bibinfo{person}{Shaohua Jiang},
  \bibinfo{person}{Yuan Qi}, {and} \bibinfo{person}{Le Song}.}
  \bibinfo{year}{2019}\natexlab{b}.
\newblock \showarticletitle{{Generative Adversarial User Model for
  Reinforcement Learning based Recommendation System}}. In
  \bibinfo{booktitle}{\emph{International Conference on Machine Learning}}.
  PMLR, \bibinfo{pages}{1052--1061}.
\newblock


\bibitem[Cheng et~al\mbox{.}(2016)]%
        {cheng2016wide}
\bibfield{author}{\bibinfo{person}{Heng-Tze Cheng}, \bibinfo{person}{Levent
  Koc}, \bibinfo{person}{Jeremiah Harmsen}, \bibinfo{person}{Tal Shaked},
  \bibinfo{person}{Tushar Chandra}, \bibinfo{person}{Hrishi Aradhye},
  \bibinfo{person}{Glen Anderson}, \bibinfo{person}{Greg Corrado},
  \bibinfo{person}{Wei Chai}, \bibinfo{person}{Mustafa Ispir}, {et~al\mbox{.}}}
  \bibinfo{year}{2016}\natexlab{}.
\newblock \showarticletitle{{Wide \& Deep Learning for Recommender Systems}}.
  In \bibinfo{booktitle}{\emph{Proceedings of the 1st workshop on deep learning
  for recommender systems}}. \bibinfo{pages}{7--10}.
\newblock


\bibitem[Chouldechova and Roth(2018)]%
        {chouldechova2018frontiers}
\bibfield{author}{\bibinfo{person}{Alexandra Chouldechova} {and}
  \bibinfo{person}{Aaron Roth}.} \bibinfo{year}{2018}\natexlab{}.
\newblock \showarticletitle{{The Frontiers of Fairness in Machine Learning}}.
\newblock \bibinfo{journal}{\emph{arXiv preprint arXiv:1810.08810}}
  (\bibinfo{year}{2018}).
\newblock


\bibitem[Corbett-Davies and Goel(2018)]%
        {corbett2018measure}
\bibfield{author}{\bibinfo{person}{Sam Corbett-Davies} {and}
  \bibinfo{person}{Sharad Goel}.} \bibinfo{year}{2018}\natexlab{}.
\newblock \showarticletitle{{The Measure and Mismeasure of Fairness: A Critical
  Review of Fair Machine Learning}}.
\newblock \bibinfo{journal}{\emph{arXiv preprint arXiv:1808.00023}}
  (\bibinfo{year}{2018}).
\newblock


\bibitem[Desirena et~al\mbox{.}(2019)]%
        {desirena2019maximizing}
\bibfield{author}{\bibinfo{person}{Gaddiel Desirena}, \bibinfo{person}{Armando
  Diaz}, \bibinfo{person}{Jalil Desirena}, \bibinfo{person}{Ismael Moreno},
  {and} \bibinfo{person}{Daniel Garcia}.} \bibinfo{year}{2019}\natexlab{}.
\newblock \showarticletitle{Maximizing {C}ustomer {L}ifetime {V}alue {u}sing
  {S}tacked {N}eural {N}etworks: {A}n {I}nsurance {I}ndustry {A}pplication}. In
  \bibinfo{booktitle}{\emph{2019 18th IEEE International Conference On Machine
  Learning And Applications (ICMLA)}}. IEEE, \bibinfo{pages}{541--544}.
\newblock


\bibitem[Ganganwar(2012)]%
        {ganganwar2012overview}
\bibfield{author}{\bibinfo{person}{Vaishali Ganganwar}.}
  \bibinfo{year}{2012}\natexlab{}.
\newblock \showarticletitle{An {O}verview of {C}lassification {A}lgorithms for
  {I}mbalanced {D}atasets}.
\newblock \bibinfo{journal}{\emph{International Journal of Emerging Technology
  and Advanced Engineering}} \bibinfo{volume}{2}, \bibinfo{number}{4}
  (\bibinfo{year}{2012}), \bibinfo{pages}{42--47}.
\newblock


\bibitem[Garivier and Capp{\'e}(2011)]%
        {garivier2011kl}
\bibfield{author}{\bibinfo{person}{Aur{\'e}lien Garivier} {and}
  \bibinfo{person}{Olivier Capp{\'e}}.} \bibinfo{year}{2011}\natexlab{}.
\newblock \showarticletitle{{The KL-UCB Algorithm for Bounded Stochastic
  Bandits and Beyond}}. In \bibinfo{booktitle}{\emph{Proceedings of the 24th
  annual conference on learning theory}}. JMLR Workshop and Conference
  Proceedings, \bibinfo{pages}{359--376}.
\newblock


\bibitem[Glorot and Bengio(2010)]%
        {glorot2010understanding}
\bibfield{author}{\bibinfo{person}{Xavier Glorot} {and} \bibinfo{person}{Yoshua
  Bengio}.} \bibinfo{year}{2010}\natexlab{}.
\newblock \showarticletitle{{Understanding the Difficulty of Training Deep
  Feedforward Neural Networks}}. In \bibinfo{booktitle}{\emph{Proceedings of
  the Thirteenth International Conference on Artificial Intelligence and
  Statistics}}. \bibinfo{pages}{249--256}.
\newblock


\bibitem[Golovin and Rahm(2004)]%
        {golovin2004reinforcement}
\bibfield{author}{\bibinfo{person}{Nick Golovin} {and} \bibinfo{person}{Erhard
  Rahm}.} \bibinfo{year}{2004}\natexlab{}.
\newblock \showarticletitle{{Reinforcement Learning Architecture for Web
  Recommendations}}. In \bibinfo{booktitle}{\emph{International Conference on
  Information Technology: Coding and Computing, 2004. Proceedings. ITCC
  2004.}}, Vol.~\bibinfo{volume}{1}. IEEE, \bibinfo{pages}{398--402}.
\newblock


\bibitem[Goodfellow et~al\mbox{.}(2020)]%
        {goodfellow2020generative}
\bibfield{author}{\bibinfo{person}{Ian Goodfellow}, \bibinfo{person}{Jean
  Pouget-Abadie}, \bibinfo{person}{Mehdi Mirza}, \bibinfo{person}{Bing Xu},
  \bibinfo{person}{David Warde-Farley}, \bibinfo{person}{Sherjil Ozair},
  \bibinfo{person}{Aaron Courville}, {and} \bibinfo{person}{Yoshua Bengio}.}
  \bibinfo{year}{2020}\natexlab{}.
\newblock \showarticletitle{{Generative Adversarial Networks}}.
\newblock \bibinfo{journal}{\emph{Commun. ACM}} \bibinfo{volume}{63},
  \bibinfo{number}{11} (\bibinfo{year}{2020}), \bibinfo{pages}{139--144}.
\newblock


\bibitem[Guo et~al\mbox{.}(2023)]%
        {Guo2023evaluate}
\bibfield{author}{\bibinfo{person}{Hongbo Guo}, \bibinfo{person}{Ruben Naeff},
  \bibinfo{person}{Alex Nikulkov}, {and} \bibinfo{person}{Zheqing Zhu}.}
  \bibinfo{year}{2023}\natexlab{}.
\newblock \showarticletitle{Evaluating Online Bandit Exploration In Large-Scale
  Recommender System}. In \bibinfo{booktitle}{\emph{KDD-23 Workshop on
  Multi-Armed Bandits and Reinforcement Learning: Advancing Decision Making in
  E-Commerce and Beyond}}.
\newblock


\bibitem[He et~al\mbox{.}(2018)]%
        {he2017neural}
\bibfield{author}{\bibinfo{person}{Xiangnan He}, \bibinfo{person}{Xiaoyu Du},
  \bibinfo{person}{Xiang Wang}, \bibinfo{person}{Feng Tian},
  \bibinfo{person}{Jinhui Tang}, {and} \bibinfo{person}{Tat-Seng Chua}.}
  \bibinfo{year}{2018}\natexlab{}.
\newblock \showarticletitle{{Outer Product-based Neural Collaborative
  Filtering}}. In \bibinfo{booktitle}{\emph{Proceedings of the Twenty-Seventh
  International Joint Conference on Artificial Intelligence}}.
  \bibinfo{pages}{2227--2233}.
\newblock


\bibitem[Heyse et~al\mbox{.}(2019)]%
        {heyse2019contextual}
\bibfield{author}{\bibinfo{person}{Joris Heyse}, \bibinfo{person}{Maria~Torres
  Vega}, \bibinfo{person}{Femke De~Backere}, {and} \bibinfo{person}{Filip
  De~Turck}.} \bibinfo{year}{2019}\natexlab{}.
\newblock \showarticletitle{{Contextual Bandit Learning-Based Viewport
  Prediction for 360 Video}}. In \bibinfo{booktitle}{\emph{2019 IEEE Conference
  on Virtual Reality and 3D User Interfaces (VR)}}. IEEE,
  \bibinfo{pages}{972--973}.
\newblock


\bibitem[Hu et~al\mbox{.}(2017)]%
        {hu2017playlist}
\bibfield{author}{\bibinfo{person}{Binbin Hu}, \bibinfo{person}{Chuan Shi},
  {and} \bibinfo{person}{Jian Liu}.} \bibinfo{year}{2017}\natexlab{}.
\newblock \showarticletitle{{Playlist Recommendation Based on Reinforcement
  Learning}}. In \bibinfo{booktitle}{\emph{International Conference on
  Intelligence Science}}. Springer, \bibinfo{pages}{172--182}.
\newblock


\bibitem[Ie et~al\mbox{.}(2019)]%
        {eugene2019slateq}
\bibfield{author}{\bibinfo{person}{Eugene Ie}, \bibinfo{person}{Vihan Jain},
  \bibinfo{person}{Jing Wang}, \bibinfo{person}{Sanmit Narvekar},
  \bibinfo{person}{Ritesh Agarwal}, \bibinfo{person}{Rui Wu},
  \bibinfo{person}{Heng-Tze Cheng}, \bibinfo{person}{Tushar Chandra}, {and}
  \bibinfo{person}{Craig Boutilier}.} \bibinfo{year}{2019}\natexlab{}.
\newblock \showarticletitle{Slate{Q}: A {T}ractable {D}ecomposition for
  {R}einforcement {L}earning with {R}ecommendation {S}ets}. In
  \bibinfo{booktitle}{\emph{Proceedings of the Twenty-eighth International
  Joint Conference on Artificial Intelligence (IJCAI-19)}}.
  \bibinfo{address}{Macau, China}, \bibinfo{pages}{2592--2599}.
\newblock
\newblock
\shownote{See arXiv:1905.12767 for a related and expanded paper (with
  additional material and authors).}.


\bibitem[Iwata et~al\mbox{.}(2008)]%
        {iwata2008recommendation}
\bibfield{author}{\bibinfo{person}{Tomoharu Iwata}, \bibinfo{person}{Kazumi
  Saito}, {and} \bibinfo{person}{Takeshi Yamada}.}
  \bibinfo{year}{2008}\natexlab{}.
\newblock \showarticletitle{Recommendation {M}ethod for {I}mproving {C}ustomer
  {L}ifetime {V}alue}.
\newblock \bibinfo{journal}{\emph{IEEE Transactions on Knowledge and Data
  Engineering}} \bibinfo{volume}{20}, \bibinfo{number}{9}
  (\bibinfo{year}{2008}), \bibinfo{pages}{1254--1263}.
\newblock


\bibitem[Joulani et~al\mbox{.}(2013)]%
        {joulani2013online}
\bibfield{author}{\bibinfo{person}{Pooria Joulani}, \bibinfo{person}{Andras
  Gyorgy}, {and} \bibinfo{person}{Csaba Szepesv{\'a}ri}.}
  \bibinfo{year}{2013}\natexlab{}.
\newblock \showarticletitle{Online {L}earning under {D}elayed {F}eedback}. In
  \bibinfo{booktitle}{\emph{International Conference on Machine Learning}}.
  \bibinfo{pages}{1453--1461}.
\newblock


\bibitem[Kakade(2003)]%
        {kakade2003sample}
\bibfield{author}{\bibinfo{person}{Sham~Machandranath Kakade}.}
  \bibinfo{year}{2003}\natexlab{}.
\newblock \bibinfo{booktitle}{\emph{{On the Sample Complexity of Reinforcement
  Learning}}}.
\newblock \bibinfo{publisher}{University of London, University College London
  (United Kingdom)}.
\newblock


\bibitem[Kamishima and Akaho(2011)]%
        {kamishima2011personalized}
\bibfield{author}{\bibinfo{person}{Toshihiro Kamishima} {and}
  \bibinfo{person}{Shotaro Akaho}.} \bibinfo{year}{2011}\natexlab{}.
\newblock \showarticletitle{{Personalized Pricing Recommender System:
  Multi-Stage Epsilon-Greedy Approach}}. In
  \bibinfo{booktitle}{\emph{Proceedings of the 2nd International Workshop on
  Information Heterogeneity and Fusion in Recommender Systems}}.
  \bibinfo{pages}{57--64}.
\newblock


\bibitem[Katehakis and Veinott~Jr(1987)]%
        {katehakis1987multi}
\bibfield{author}{\bibinfo{person}{Michael~N Katehakis} {and}
  \bibinfo{person}{Arthur~F Veinott~Jr}.} \bibinfo{year}{1987}\natexlab{}.
\newblock \showarticletitle{{The Multi-Armed Bandit Problem: Decomposition and
  Computation}}.
\newblock \bibinfo{journal}{\emph{Mathematics of Operations Research}}
  \bibinfo{volume}{12}, \bibinfo{number}{2} (\bibinfo{year}{1987}),
  \bibinfo{pages}{262--268}.
\newblock


\bibitem[Ktena et~al\mbox{.}(2019)]%
        {ktena2019addressing}
\bibfield{author}{\bibinfo{person}{Sofia~Ira Ktena}, \bibinfo{person}{Alykhan
  Tejani}, \bibinfo{person}{Lucas Theis}, \bibinfo{person}{Pranay~Kumar Myana},
  \bibinfo{person}{Deepak Dilipkumar}, \bibinfo{person}{Ferenc Husz{\'a}r},
  \bibinfo{person}{Steven Yoo}, {and} \bibinfo{person}{Wenzhe Shi}.}
  \bibinfo{year}{2019}\natexlab{}.
\newblock \showarticletitle{Addressing {D}elayed {F}eedback for {C}ontinuous
  {T}raining with {N}eural {N}etworks in {C}TR {P}rediction}. In
  \bibinfo{booktitle}{\emph{Proceedings of the 13th ACM Conference on
  Recommender Systems}}. \bibinfo{pages}{187--195}.
\newblock


\bibitem[Langford and Zhang(2007)]%
        {langford2007epoch}
\bibfield{author}{\bibinfo{person}{John Langford} {and} \bibinfo{person}{Tong
  Zhang}.} \bibinfo{year}{2007}\natexlab{}.
\newblock \showarticletitle{{Epoch-Greedy Algorithm for Multi-Armed Bandits
  with Side Information}}.
\newblock \bibinfo{journal}{\emph{Advances in Neural Information Processing
  Systems (NIPS 2007)}}  \bibinfo{volume}{20} (\bibinfo{year}{2007}),
  \bibinfo{pages}{1}.
\newblock


\bibitem[Li et~al\mbox{.}(2010)]%
        {li2010contextual}
\bibfield{author}{\bibinfo{person}{Lihong Li}, \bibinfo{person}{Wei Chu},
  \bibinfo{person}{John Langford}, {and} \bibinfo{person}{Robert~E Schapire}.}
  \bibinfo{year}{2010}\natexlab{}.
\newblock \showarticletitle{{A Contextual-Bandit Approach to Personalized News
  Article Recommendation}}. In \bibinfo{booktitle}{\emph{Proceedings of the
  19th International Conference on World Wide Web}}. \bibinfo{pages}{661--670}.
\newblock


\bibitem[Li et~al\mbox{.}(2011)]%
        {li2011unbiased}
\bibfield{author}{\bibinfo{person}{Lihong Li}, \bibinfo{person}{Wei Chu},
  \bibinfo{person}{John Langford}, {and} \bibinfo{person}{Xuanhui Wang}.}
  \bibinfo{year}{2011}\natexlab{}.
\newblock \showarticletitle{{Unbiased Offline Evaluation of
  Contextual-Bandit-Based News Article Recommendation Algorithms}}. In
  \bibinfo{booktitle}{\emph{Proceedings of the Fourth ACM International
  Conference on Web Search and Data Mining}}. ACM, \bibinfo{pages}{297--306}.
\newblock


\bibitem[Liu and Shih(2005)]%
        {liu2005integrating}
\bibfield{author}{\bibinfo{person}{Duen-Ren Liu} {and} \bibinfo{person}{Ya-Yueh
  Shih}.} \bibinfo{year}{2005}\natexlab{}.
\newblock \showarticletitle{Integrating {A}HP and {D}ata {M}ining for {P}roduct
  {R}ecommendation {B}ased on {C}ustomer {L}ifetime {V}salue}.
\newblock \bibinfo{journal}{\emph{Information \& Management}}
  \bibinfo{volume}{42}, \bibinfo{number}{3} (\bibinfo{year}{2005}),
  \bibinfo{pages}{387--400}.
\newblock


\bibitem[Lu and Van~Roy(2017)]%
        {lu2017ensemble}
\bibfield{author}{\bibinfo{person}{Xiuyuan Lu} {and} \bibinfo{person}{Benjamin
  Van~Roy}.} \bibinfo{year}{2017}\natexlab{}.
\newblock \showarticletitle{{Ensemble Sampling}}. In
  \bibinfo{booktitle}{\emph{Advances in Neural Information Processing
  Systems}}. \bibinfo{pages}{3258--3266}.
\newblock


\bibitem[Mnih et~al\mbox{.}(2015)]%
        {mnih2015human}
\bibfield{author}{\bibinfo{person}{Volodymyr Mnih}, \bibinfo{person}{Koray
  Kavukcuoglu}, \bibinfo{person}{David Silver}, \bibinfo{person}{Andrei~A
  Rusu}, \bibinfo{person}{Joel Veness}, \bibinfo{person}{Marc~G Bellemare},
  \bibinfo{person}{Alex Graves}, \bibinfo{person}{Martin Riedmiller},
  \bibinfo{person}{Andreas~K Fidjeland}, \bibinfo{person}{Georg Ostrovski},
  {et~al\mbox{.}}} \bibinfo{year}{2015}\natexlab{}.
\newblock \showarticletitle{{Human-Level Control through Deep Reinforcement
  Learning}}.
\newblock \bibinfo{journal}{\emph{Nature}} \bibinfo{volume}{518},
  \bibinfo{number}{7540} (\bibinfo{year}{2015}), \bibinfo{pages}{529--533}.
\newblock


\bibitem[Nakamura(2015)]%
        {nakamura2015ucb}
\bibfield{author}{\bibinfo{person}{Atsuyoshi Nakamura}.}
  \bibinfo{year}{2015}\natexlab{}.
\newblock \showarticletitle{{A UCB-Like Strategy of Collaborative Filtering}}.
  In \bibinfo{booktitle}{\emph{Asian Conference on Machine Learning}}.
  \bibinfo{pages}{315--329}.
\newblock


\bibitem[Nguyen-Thanh et~al\mbox{.}(2019)]%
        {nguyen2019recommendation}
\bibfield{author}{\bibinfo{person}{Nhan Nguyen-Thanh}, \bibinfo{person}{Dana
  Marinca}, \bibinfo{person}{Kinda Khawam}, \bibinfo{person}{David Rohde},
  \bibinfo{person}{Flavian Vasile}, \bibinfo{person}{Elena Lohan},
  \bibinfo{person}{Steven Martin}, {and} \bibinfo{person}{Dominique Quadri}.}
  \bibinfo{year}{2019}\natexlab{}.
\newblock \showarticletitle{Recommendation {S}ystem-{B}ased {U}pper
  {C}onfidence {B}ound for {O}nline {A}dvertising}. In
  \bibinfo{booktitle}{\emph{REVEAL 2019}}.
\newblock


\bibitem[Osband et~al\mbox{.}(2018)]%
        {osband2018randomized}
\bibfield{author}{\bibinfo{person}{Ian Osband}, \bibinfo{person}{John
  Aslanides}, {and} \bibinfo{person}{Albin Cassirer}.}
  \bibinfo{year}{2018}\natexlab{}.
\newblock \showarticletitle{{Randomized Prior Functions for Deep Reinforcement
  Learning}}. In \bibinfo{booktitle}{\emph{Advances in Neural Information
  Processing Systems}}. \bibinfo{pages}{8617--8629}.
\newblock


\bibitem[Osband et~al\mbox{.}(2016)]%
        {osband2016deep}
\bibfield{author}{\bibinfo{person}{Ian Osband}, \bibinfo{person}{Charles
  Blundell}, \bibinfo{person}{Alexander Pritzel}, {and}
  \bibinfo{person}{Benjamin Van~Roy}.} \bibinfo{year}{2016}\natexlab{}.
\newblock \showarticletitle{{Deep Exploration via Bootstrapped DQN}}. In
  \bibinfo{booktitle}{\emph{Advances in Neural information processing
  systems}}.
\newblock


\bibitem[Osband et~al\mbox{.}(2019)]%
        {osband2019deep}
\bibfield{author}{\bibinfo{person}{Ian Osband}, \bibinfo{person}{Benjamin
  Van~Roy}, \bibinfo{person}{Daniel~J Russo}, {and} \bibinfo{person}{Zheng
  Wen}.} \bibinfo{year}{2019}\natexlab{}.
\newblock \showarticletitle{{Deep Exploration via Randomized Value Functions}}.
\newblock \bibinfo{journal}{\emph{Journal of Machine Learning Research}}
  \bibinfo{volume}{20}, \bibinfo{number}{124} (\bibinfo{year}{2019}),
  \bibinfo{pages}{1--62}.
\newblock


\bibitem[Osband et~al\mbox{.}(2021)]%
        {osband2021epistemic}
\bibfield{author}{\bibinfo{person}{Ian Osband}, \bibinfo{person}{Zheng Wen},
  \bibinfo{person}{Mohammad Asghari}, \bibinfo{person}{Morteza Ibrahimi},
  \bibinfo{person}{Xiyuan Lu}, {and} \bibinfo{person}{Benjamin Van~Roy}.}
  \bibinfo{year}{2021}\natexlab{}.
\newblock \showarticletitle{{Epistemic Neural Networks}}.
\newblock \bibinfo{journal}{\emph{arXiv preprint arXiv:2107.08924}}
  (\bibinfo{year}{2021}).
\newblock


\bibitem[Osband et~al\mbox{.}(2022)]%
        {OsbandNeuralTestbed}
\bibfield{author}{\bibinfo{person}{Ian Osband}, \bibinfo{person}{Zhegn Wen},
  \bibinfo{person}{Seyed~Mohammad Asghari}, \bibinfo{person}{Vikranth
  Dwaracherla}, \bibinfo{person}{Xiuyuan Lu}, \bibinfo{person}{Morteza
  Ibrahimi}, \bibinfo{person}{Dieterich Lawson}, \bibinfo{person}{Botao Hao},
  \bibinfo{person}{Brendan O'Donoghue}, {and} \bibinfo{person}{Benjamin
  Van~Roy}.} \bibinfo{year}{2022}\natexlab{}.
\newblock \showarticletitle{{The Neural Testbed: Evaluating Joint
  Predictions}}. In \bibinfo{booktitle}{\emph{Advances in Neural Information
  Processing Systems}}.
\newblock


\bibitem[Qin et~al\mbox{.}(2022)]%
        {QinEnsembleSampling}
\bibfield{author}{\bibinfo{person}{Chao Qin}, \bibinfo{person}{Zheng Wen},
  \bibinfo{person}{Xiuyuan Lu}, {and} \bibinfo{person}{Benjamin Van~Roy}.}
  \bibinfo{year}{2022}\natexlab{}.
\newblock \showarticletitle{{An Analysis of Ensemble Sampling}}. In
  \bibinfo{booktitle}{\emph{Advances in Neural Information Processing
  Systems}}.
\newblock


\bibitem[Rojanavasu et~al\mbox{.}(2005)]%
        {rojanavasu2005new}
\bibfield{author}{\bibinfo{person}{Pornthep Rojanavasu},
  \bibinfo{person}{Phaitoon Srinil}, {and} \bibinfo{person}{Ouen Pinngern}.}
  \bibinfo{year}{2005}\natexlab{}.
\newblock \showarticletitle{{New Recommendation System Using Reinforcement
  Learning}}.
\newblock \bibinfo{journal}{\emph{Special Issue of the Intl. J. Computer, the
  Internet and Management}} \bibinfo{volume}{13}, \bibinfo{number}{SP 3}
  (\bibinfo{year}{2005}).
\newblock


\bibitem[Russo and Van~Roy(2014)]%
        {russo2014learning}
\bibfield{author}{\bibinfo{person}{Daniel Russo} {and}
  \bibinfo{person}{Benjamin Van~Roy}.} \bibinfo{year}{2014}\natexlab{}.
\newblock \showarticletitle{{Learning to Optimize via Information-Directed
  Sampling}}. In \bibinfo{booktitle}{\emph{Advances in Neural Information
  Processing Systems}}. \bibinfo{pages}{1583--1591}.
\newblock


\bibitem[Schafer et~al\mbox{.}(2007)]%
        {schafer2007collaborative}
\bibfield{author}{\bibinfo{person}{J~Ben Schafer}, \bibinfo{person}{Dan
  Frankowski}, \bibinfo{person}{Jon Herlocker}, {and} \bibinfo{person}{Shilad
  Sen}.} \bibinfo{year}{2007}\natexlab{}.
\newblock \showarticletitle{Collaborative {F}iltering {R}ecommender {S}ystems}.
\newblock \bibinfo{journal}{\emph{The Adaptive Web}} (\bibinfo{year}{2007}),
  \bibinfo{pages}{291--324}.
\newblock


\bibitem[Schrittwieser et~al\mbox{.}(2020)]%
        {schrittwieser2020mastering}
\bibfield{author}{\bibinfo{person}{Julian Schrittwieser},
  \bibinfo{person}{Ioannis Antonoglou}, \bibinfo{person}{Thomas Hubert},
  \bibinfo{person}{Karen Simonyan}, \bibinfo{person}{Laurent Sifre},
  \bibinfo{person}{Simon Schmitt}, \bibinfo{person}{Arthur Guez},
  \bibinfo{person}{Edward Lockhart}, \bibinfo{person}{Demis Hassabis},
  \bibinfo{person}{Thore Graepel}, {et~al\mbox{.}}}
  \bibinfo{year}{2020}\natexlab{}.
\newblock \showarticletitle{{Mastering Atari, Go, Chess and Shogi by Planning
  with a Learned Model}}.
\newblock \bibinfo{journal}{\emph{Nature}} \bibinfo{volume}{588},
  \bibinfo{number}{7839} (\bibinfo{year}{2020}), \bibinfo{pages}{604--609}.
\newblock


\bibitem[Shi et~al\mbox{.}(2019)]%
        {shi2019virtual}
\bibfield{author}{\bibinfo{person}{Jing-Cheng Shi}, \bibinfo{person}{Yang Yu},
  \bibinfo{person}{Qing Da}, \bibinfo{person}{Shi-Yong Chen}, {and}
  \bibinfo{person}{An-Xiang Zeng}.} \bibinfo{year}{2019}\natexlab{}.
\newblock \showarticletitle{{Virtual-Taobao: Virtualizing Real-World Online
  Retail Environment for Reinforcement Learning}}. In
  \bibinfo{booktitle}{\emph{Proceedings of the AAAI Conference on Artificial
  Intelligence}}, Vol.~\bibinfo{volume}{33}. \bibinfo{pages}{4902--4909}.
\newblock


\bibitem[Shih and Liu(2008)]%
        {shih2008product}
\bibfield{author}{\bibinfo{person}{Ya-Yueh Shih} {and}
  \bibinfo{person}{Duen-Ren Liu}.} \bibinfo{year}{2008}\natexlab{}.
\newblock \showarticletitle{Product {R}ecommendation {A}pproaches:
  {C}ollaborative {F}iltering via {C}ustomer {L}ifetime {V}alue and {C}ustomer
  {D}emands}.
\newblock \bibinfo{journal}{\emph{Expert Systems with Applications}}
  \bibinfo{volume}{35}, \bibinfo{number}{1-2} (\bibinfo{year}{2008}),
  \bibinfo{pages}{350--360}.
\newblock


\bibitem[Silver et~al\mbox{.}(2017)]%
        {silver2017mastering}
\bibfield{author}{\bibinfo{person}{David Silver}, \bibinfo{person}{Julian
  Schrittwieser}, \bibinfo{person}{Karen Simonyan}, \bibinfo{person}{Ioannis
  Antonoglou}, \bibinfo{person}{Aja Huang}, \bibinfo{person}{Arthur Guez},
  \bibinfo{person}{Thomas Hubert}, \bibinfo{person}{Lucas Baker},
  \bibinfo{person}{Matthew Lai}, \bibinfo{person}{Adrian Bolton},
  {et~al\mbox{.}}} \bibinfo{year}{2017}\natexlab{}.
\newblock \showarticletitle{{Mastering the Game of Go without Human
  Knowledge}}.
\newblock \bibinfo{journal}{\emph{nature}} \bibinfo{volume}{550},
  \bibinfo{number}{7676} (\bibinfo{year}{2017}), \bibinfo{pages}{354--359}.
\newblock


\bibitem[Song et~al\mbox{.}(2018)]%
        {song2018multi}
\bibfield{author}{\bibinfo{person}{Jiaming Song}, \bibinfo{person}{Hongyu Ren},
  \bibinfo{person}{Dorsa Sadigh}, {and} \bibinfo{person}{Stefano Ermon}.}
  \bibinfo{year}{2018}\natexlab{}.
\newblock \showarticletitle{{Multi-Agent Generative Adversarial Imitation
  Learning}}. In \bibinfo{booktitle}{\emph{Advances in neural information
  processing systems}}.
\newblock


\bibitem[Tang et~al\mbox{.}(2019)]%
        {tang2019reinforcement}
\bibfield{author}{\bibinfo{person}{Xueying Tang}, \bibinfo{person}{Yunxiao
  Chen}, \bibinfo{person}{Xiaoou Li}, \bibinfo{person}{Jingchen Liu}, {and}
  \bibinfo{person}{Zhiliang Ying}.} \bibinfo{year}{2019}\natexlab{}.
\newblock \showarticletitle{{A Reinforcement Learning Approach to Personalized
  Learning Recommendation Systems}}.
\newblock \bibinfo{journal}{\emph{Brit. J. Math. Statist. Psych.}}
  \bibinfo{volume}{72}, \bibinfo{number}{1} (\bibinfo{year}{2019}),
  \bibinfo{pages}{108--135}.
\newblock


\bibitem[Theocharous et~al\mbox{.}(2015)]%
        {theocharous2015personalized}
\bibfield{author}{\bibinfo{person}{Georgios Theocharous},
  \bibinfo{person}{Philip~S Thomas}, {and} \bibinfo{person}{Mohammad
  Ghavamzadeh}.} \bibinfo{year}{2015}\natexlab{}.
\newblock \showarticletitle{Personalized {A}d {R}ecommendation {S}ystems for
  {L}ife-{T}ime {V}alue {O}ptimization with {G}uarantees}. In
  \bibinfo{booktitle}{\emph{Twenty-Fourth International Joint Conference on
  Artificial Intelligence}}.
\newblock


\bibitem[Thompson(1933)]%
        {thompson1933likelihood}
\bibfield{author}{\bibinfo{person}{William~R Thompson}.}
  \bibinfo{year}{1933}\natexlab{}.
\newblock \showarticletitle{{On the Likelihood That One Unknown Probability
  Exceeds Another in View of the Evidence of Two Samples}}.
\newblock \bibinfo{journal}{\emph{Biometrika}} \bibinfo{volume}{25},
  \bibinfo{number}{3/4} (\bibinfo{year}{1933}), \bibinfo{pages}{285--294}.
\newblock


\bibitem[Wang et~al\mbox{.}(2021)]%
        {wang2021rl4rs}
\bibfield{author}{\bibinfo{person}{Kai Wang}, \bibinfo{person}{Zhene Zou},
  \bibinfo{person}{Qilin Deng}, \bibinfo{person}{Yue Shang},
  \bibinfo{person}{Minghao Zhao}, \bibinfo{person}{Runze Wu},
  \bibinfo{person}{Xudong Shen}, \bibinfo{person}{Tangjie Lyu}, {and}
  \bibinfo{person}{Changjie Fan}.} \bibinfo{year}{2021}\natexlab{}.
\newblock \showarticletitle{{RL4RS: A Real-World Benchmark for Reinforcement
  Learning based Recommender System}}.
\newblock \bibinfo{journal}{\emph{arXiv preprint arXiv:2110.11073}}
  (\bibinfo{year}{2021}).
\newblock


\bibitem[Wu et~al\mbox{.}(2019)]%
        {wu2019neural}
\bibfield{author}{\bibinfo{person}{Chuhan Wu}, \bibinfo{person}{Fangzhao Wu},
  \bibinfo{person}{Suyu Ge}, \bibinfo{person}{Tao Qi},
  \bibinfo{person}{Yongfeng Huang}, {and} \bibinfo{person}{Xing Xie}.}
  \bibinfo{year}{2019}\natexlab{}.
\newblock \showarticletitle{{Neural News Recommendation with Multi-Head
  Self-Attention}}. In \bibinfo{booktitle}{\emph{Proceedings of the 2019
  conference on empirical methods in natural language processing and the 9th
  international joint conference on natural language processing
  (EMNLP-IJCNLP)}}. \bibinfo{pages}{6389--6394}.
\newblock


\bibitem[Xu et~al\mbox{.}(2021)]%
        {xu2021neural}
\bibfield{author}{\bibinfo{person}{Pan Xu}, \bibinfo{person}{Zheng Wen},
  \bibinfo{person}{Handong Zhao}, {and} \bibinfo{person}{Quanquan Gu}.}
  \bibinfo{year}{2021}\natexlab{}.
\newblock \showarticletitle{{Neural Contextual Bandits with Deep Representation
  and Shallow Exploration}}. In \bibinfo{booktitle}{\emph{International
  Conference on Learning Representations}}.
\newblock


\bibitem[Xu et~al\mbox{.}(2023)]%
        {xu2023optimize}
\bibfield{author}{\bibinfo{person}{Ruiyang Xu}, \bibinfo{person}{Jalaj
  Bhandari}, \bibinfo{person}{Dmytro Korenkevych}, \bibinfo{person}{Fan Liu},
  \bibinfo{person}{Yuchen He}, \bibinfo{person}{Alex Nikulkov}, {and}
  \bibinfo{person}{Zheqing Zhu}.} \bibinfo{year}{2023}\natexlab{}.
\newblock \showarticletitle{Optimizing Long-term Value for Auction-Based
  Recommender Systems via On-Policy Reinforcement Learning}. In
  \bibinfo{booktitle}{\emph{Proceedings of the 17th ACM Conference on
  Recommender Systems}}.
\newblock


\bibitem[Zhang et~al\mbox{.}(2017)]%
        {zhang2017understanding}
\bibfield{author}{\bibinfo{person}{C Zhang}, \bibinfo{person}{S Bengio},
  \bibinfo{person}{M Hardt}, \bibinfo{person}{B Recht}, {and}
  \bibinfo{person}{O Vinyals}.} \bibinfo{year}{2017}\natexlab{}.
\newblock \showarticletitle{Understanding Deep Learning Requires Rethinking
  Generalization Int}. In \bibinfo{booktitle}{\emph{Conf. on Learning
  Representations}}.
\newblock


\bibitem[Zhang et~al\mbox{.}(2020)]%
        {zhang2020neural}
\bibfield{author}{\bibinfo{person}{Weitong Zhang}, \bibinfo{person}{Dongruo
  Zhou}, \bibinfo{person}{Lihong Li}, {and} \bibinfo{person}{Quanquan Gu}.}
  \bibinfo{year}{2020}\natexlab{}.
\newblock \showarticletitle{{Neural Thompson Sampling}}. In
  \bibinfo{booktitle}{\emph{International Conference on Learning
  Representations}}.
\newblock


\bibitem[Zheng et~al\mbox{.}(2018)]%
        {zheng2018drn}
\bibfield{author}{\bibinfo{person}{Guanjie Zheng}, \bibinfo{person}{Fuzheng
  Zhang}, \bibinfo{person}{Zihan Zheng}, \bibinfo{person}{Yang Xiang},
  \bibinfo{person}{Nicholas~Jing Yuan}, \bibinfo{person}{Xing Xie}, {and}
  \bibinfo{person}{Zhenhui Li}.} \bibinfo{year}{2018}\natexlab{}.
\newblock \showarticletitle{{DRN}: A {D}eep {R}einforcement {L}earning
  {F}ramework for {N}ews {R}ecommendation}. In
  \bibinfo{booktitle}{\emph{Proceedings of the 2018 World Wide Web Conference
  on World Wide Web}}. International World Wide Web Conferences Steering
  Committee, \bibinfo{pages}{167--176}.
\newblock


\bibitem[Zhou et~al\mbox{.}(2020)]%
        {zhou2019neural}
\bibfield{author}{\bibinfo{person}{Dongruo Zhou}, \bibinfo{person}{Lihong Li},
  {and} \bibinfo{person}{Quanquan Gu}.} \bibinfo{year}{2020}\natexlab{}.
\newblock \showarticletitle{{Neural Contextual Bandits with UCB-Based
  Exploration}}. In \bibinfo{booktitle}{\emph{International Conference on
  Machine Learning}}. PMLR, \bibinfo{pages}{11492--11502}.
\newblock


\bibitem[Zhou et~al\mbox{.}(2019)]%
        {zhou2019deep}
\bibfield{author}{\bibinfo{person}{Guorui Zhou}, \bibinfo{person}{Na Mou},
  \bibinfo{person}{Ying Fan}, \bibinfo{person}{Qi Pi}, \bibinfo{person}{Weijie
  Bian}, \bibinfo{person}{Chang Zhou}, \bibinfo{person}{Xiaoqiang Zhu}, {and}
  \bibinfo{person}{Kun Gai}.} \bibinfo{year}{2019}\natexlab{}.
\newblock \showarticletitle{{Deep Interest Evolution Network for Click-Through
  Rate Prediction}}. In \bibinfo{booktitle}{\emph{Proceedings of the AAAI
  conference on artificial intelligence}}, Vol.~\bibinfo{volume}{33}.
  \bibinfo{pages}{5941--5948}.
\newblock


\bibitem[Zhu and Van~Roy(2023)]%
        {zhu2023scalable}
\bibfield{author}{\bibinfo{person}{Zheqing Zhu} {and} \bibinfo{person}{Benjamin
  Van~Roy}.} \bibinfo{year}{2023}\natexlab{}.
\newblock \showarticletitle{Scalable Neural Contextual Bandit for Recommender
  Systems}.
\newblock \bibinfo{journal}{\emph{arXiv preprint arXiv:2306.14834}}
  (\bibinfo{year}{2023}).
\newblock


\end{thebibliography}

\end{document}